\documentclass[remotesensing,article,submit,moreauthors,pdftex]{Definitions/mdpi} 
\usepackage{graphicx,hyperref,units}

\firstpage{1} 
\pubvolume{1}
\issuenum{1}
\articlenumber{5}
\pubyear{2021}
\copyrightyear{2021}

\Title{Fusion of Rain Radar Images and Wind Forecasts in a Deep Learning Model Applied to Rain Nowcasting}

\Author{Vincent 
 Bouget $^{1}$\orcidC{}, Dominique B\'er\'eziat $^{1}$\orcidA{}, Julien Brajard $^{2}$\orcidB{}, Anastase Charantonis $^{3,4,}$*\orcidD{} and Arthur Filoche $^{1}$}

\AuthorNames{Vincent Bouget, Dominique B\'er\'eziat, Julien Brajard, Anastase Charantonis, Arthur Filoche}

\address{%
$^{1}$ \quad Sorbonne 
 Universit\'e, CNRS, Laboratoire d’Informatique de Paris 6, 75005 Paris, 
 France; dominique.bereziat@lip6.fr (D.B.); arthur.filoche@lip6.fr (A.F.)\\
$^{2}$ \quad 
Nansen Environmental and Remote Sensing Center (NERSC), 5009 Bergen, Norway; julien.brajard@nersc.no\\
$^{3}$ \quad Laboratoire d’océanographie et du climat (LOCEAN), 75005 Paris, France\\
$^{4}$ \quad École nationale supérieure d'informatique pour l'industrie et l'entreprise (ENSIIE), 91000 \'Evry, France}

\corres{Correspondence: anastase.charantonis@ensiie.fr (A.C.)
}
\abstract{
Short- or mid-term rainfall forecasting is a major task with several environmental applications such as agricultural management or flood risk monitoring. Existing data-driven approaches, especially deep learning models, have shown significant skill at this task, using only rainfall radar images as inputs. In order to determine whether using other meteorological parameters such as wind would improve forecasts, we trained a deep learning model on a fusion of rainfall radar images and wind velocity produced by a weather forecast model. The network was compared to a similar architecture trained only on radar data, to a basic persistence model and to an approach based on optical flow. Our network outperforms by 8\% the \textit{F1}-score calculated for the optical flow on moderate and higher rain events for forecasts at a horizon time of 30 min. Furthermore, it outperforms by 7\% the same architecture trained using only rainfall radar images. Merging rain and wind data has also proven to stabilize the training process and enabled significant improvement especially on the difficult-to-predict high precipitation rainfalls.
}
\keyword{radar data; rain nowcasting; deep learning} 

\def\CRF{\mathit{CRF}}
\def\U{U}
\def\V{V}
\def\M{M}
\def\lon{\mathit{lon}}
\def\lat{\mathit{lat}}
\def\C{C}
\def\th{L}
\def\P{P}
\def\T{T}
\def\ov{\eta}

\preto{\abstractkeywords}{\nolinenumbers}

\begin{document}
\maketitle

\section{Introduction}
{{Forecasting}} precipitations at the short- and mid-term horizon (also known as rain
 nowcasting) is important for real-life problems, for instance, the World Meteorological Organization recently set out concrete applications in agricultural management, aviation, or management of severe meteorological events~\cite{WTH_rain_nowcasting_applications}. Rain nowcasting requires a quick and reliable forecast of a process that is highly non-stationary at a local scale. 
Due to the strong constraints of computing time, operational short-term precipitation forecasting systems are very simple in their design. To our knowledge, there are two main types of operational approaches all based on radar imagery: 
Methods based on storm cell tracking~\cite{TITAN,SCIT,TRACE3D,CELLTRACK} try to match image structures (storm cells, obtained by thresholding) seen between two successive acquisitions. Matching criteria are based on the similarity and proximity of these structures. 
Once the correspondence and their displacement have been established, the position of these cells is extrapolated to the desired time horizon. 
The second category relies on the estimation of a dense field of apparent velocities at each pixel of the image and modeled by the optical flow~\cite{MAPLE,GANDOLF}. The forecast is also obtained by extrapolation in time and advection of the last observation with the apparent velocity field.

Over the past few years, machine learning proved to be able to address rain nowcasting and was applied in several regions~\cite{Shi2015,Shi2017, China_nowcasting,Germany_nowcasting,Spain_nowcasting, Russia_nowcasting}. 
More recently, new neural network architectures were used: in~\cite{PredNet}, a PredNet~\cite{PredNEt_original} is adapted to predict rain in the region of Kyoto. In~\cite{GoogleArticle}, a U-Net architecture~\cite{UNET} is used for rain nowcasting of low- to middle-intensity rainfalls in the region of Seattle.
The key idea in these works is to train a neural network on sequences of consecutive rain radar images in order to predict the rainfall at a subsequent time. Although rain nowcasting based on deep learning is widely used, it is driven by observed radar or satellite images. In this work, we propose an algorithm merging meteorological forecasts with observed radar data to improve these predictions.

M\'et\'eo-France (the French national weather service) recently released MeteoNet~\cite{meteonet}, a database that provides a large number of meteorological parameters on the French territory. The data available are as diverse as rainfalls (acquired by Doppler radars of the M\'et\'eo-France network), the outcomes of two meteorological models (high-scale ARPEGE and finer-scale AROME), topographical masks, and so on. The outcomes of the weather forecast model AROME include hourly forecasts of wind velocity, considering that advection is a prominent factor in precipitations evolution we chose to include wind as a significant additional predictor.

The forecasts of the neural network are based on a set of parameters weighting the features of their inputs. A training procedure adjusts the network's parameters to emphasize the weights on the features significant for the network's predictions. The deep learning model used in this work is a shallow U-Net architecture~\cite{UNET} known for its skill in image processing~\cite{de2019deep}. Moreover, this architecture is flexible enough to easily add relevant inputs, which is an interesting property for data fusion. Two networks were trained on the data of MeteoNet restricted to the region of Brest in France. Their inputs were sequences of rain radar images and wind forecasts five minutes apart over an hour, and their targets were rain radar images at the horizon of 30 min for the first neural network and 1 h for the second. 
An accurate regression of rainfall is an ill-posed problem, mainly due to issues of an imbalanced dataset, heavily skewed towards null and small values. We chose to transform the problem into a classification problem, similarly to the work in ~\cite{GoogleArticle}. This approach is relevant given the potential uses of rain nowcasting, especially in predicting flash flooding, in aviation and agriculture, where the exact measurement of rain is not as important as the reaching of a threshold~\cite{WTH_rain_nowcasting_applications}. We split the rain data into several classes depending on its precipitation rate. A major issue faced during the training is rain scarcity. Given that an overwhelming number of images corresponds to a clear sky, the training dataset is imbalanced in favor of null rainfalls which makes it quite difficult for a neural network to extract significant features during training. We present a method of data over-sampling to address this issue. 

We compared our model to the persistence model which consists of taking the last rain radar image of an input sequence as the prediction (though simplistic, this model is frequently used in rain nowcasting~\cite{GoogleArticle,Spain_nowcasting,Germany_nowcasting}) and to an operational and optical flow-based rain nowcasting system~\cite{zebiri:hal-02048500}. We also compare the neural network merging radar image and wind forecast to a similar neural network trained using only radar rain images as inputs. 

\section{Problem Statement}
\label{section_problem_statement}

\begin{figure}[H]
\includegraphics[width=12 cm]{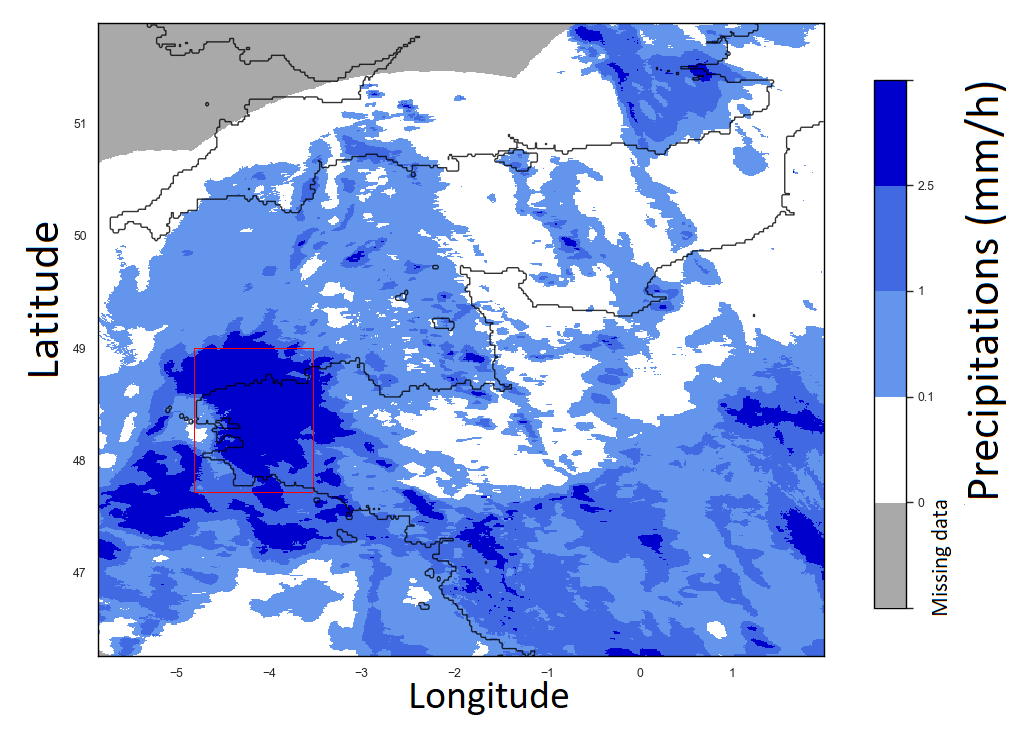}
\caption{An example of rain radar image acquired on the 2016-01-07 at 02:25:00 in the North West of France with the study area framed in red. Precipitations are colored in blue scale corresponding to the classes thresholds defined in Table~\ref{tab:Table summary of classes definition}. Gray corresponds to missing data.\label{fig:rain_sample}}
\end{figure}

\begin{figure}[H]
\includegraphics[width=12 cm]{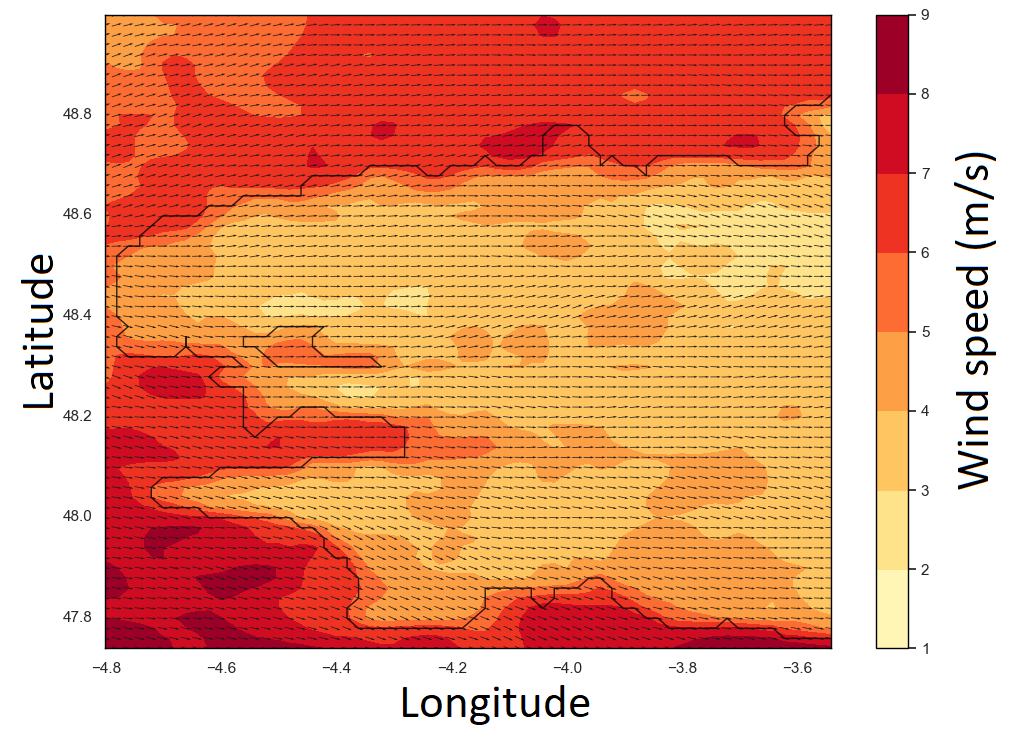}
\caption{An example of wind map restricted to the study area acquired on the 2017-05-06 at 20:25:00. Wind speed is represented by the color scale and velocity direction is indicated with arrows.}\label{fig:wind_sample}
\end{figure}

Two types of images are used: rain radar images (also referred to as rainfall maps, see Figure~\ref{fig:rain_sample}) providing for each pixel the accumulation of rainfall over 5 min and wind maps (see Figure~\ref{fig:wind_sample}) providing for each pixel the 10 m wind velocity components $U$ and $V$. Both rain and wind data will be detailed further in Section~\ref{section_data}. 

Each meteorological parameter (rainfall, wind velocity $U$, and wind velocity $V$) is available across metropolitan France at regular time steps. The images are stacked along the temporal axis.
Each pixel is indexed by three indices $(i,j,k)$; $i$ and $j$ index space and, respectively, map a data to its longitude $\lon_i$ and latitude $\lat_j$; $k$ indexes time and maps a data to its time step $t_k$. In the following, time and spatial resolutions are assumed to be constant: 0.01 degrees spatially and 5 min temporally. 

We define $\CRF_{i,j,k}$ as the cumulative rainfall between times $t_{k-1}$ and $t_k$ at longitude $\lon_i$ and latitude $\lat_j$. 
We define $\U_{i,j,k}$ and $\V_{i,j,k}$, respectively, as the horizontal (East to West) and vertical (South to North) components of the wind velocity vector at $t_k$, longitude $\lon_i$, and latitude $\lat_j$. 
Finally, we define $\M_{i,j,k} = (\CRF_{i,j,k}, U_{i,j,k}, V_{i,j,k})$ as the vector stacking all data. 
Given a sequence of MeteoNet data $\widetilde{M}_{(k-1,k-s)}=(\M_{i,j,k-1},\cdots,\M_{i,j,k-s})$ where $s \in \mathbb{N}$ is the length of the sequence, the target of our study would ideally be to forecast the rainfall at a subsequent time $\CRF_{i,j,k+p}$ where $p \in \mathbb{N}$ is the lead time step.

\unskip

As stated before, we have chosen to transform this regression problem into a classification problem. To define these classes we consider a set of ${N_L}$ ordered threshold values. 
These values split the interval $[0;+\infty)$ in $N_L$ classes defined as follows: 
for $m \in \{1,\cdots, N_L\}$, 
class $\C_m$ is defined by 
$\C_m = \{\CRF_{i,j,k} \geq \th_m \}$. 
A pixel belongs to a class if the rainfall accumulated between the time $t_{k-1}$ and the time $t_k$ is greater than the threshold associated with this class.
Splitting the cumulative rainfalls into these $N_L$ classes converts the regression task from directly predicting $\CRF_{i,j,k+p}$ to determining to which classes $\CRF_{i,j,k+p}$ belongs to.
The classes are embedded, i.e., if a $\CRF_{i,j,k}$ belongs to $\C_m$, then it also belongs to $\C_n,  \forall 1 \leq n<m$. Therefore, a prediction can belong to several classes. This type of problem with embedded classes is formalized as multi-label classification problem, and it is often transformed into $N_L$ binary classification problems using the binary relevance method~\cite{BinaryRelevanceMethod}. Therefore, we will train $N_L$ binary classifiers; classifier $m$ determines the probability that $\CRF_{i,j,k+p}$ exceeds the threshold $\th_m$. 

\nointerlineskip
\begin{table}[H]
\caption{\label{tab:Table summary of classes definition}Summary of classes definition and distribution.}
\begin{tabular}{ | c | c| c | c | c | c | } 
\hline
$\mathbf{m}$ & \textbf{Qualitative label} & \textbf{Class definition} & \textbf{Pixels by class (\%)} & \textbf{Images containing}\\
 & & & & \textbf{class m (\%)}\\ 
\hline
1 & Very light rain and higher & $\CRF \ge \unitfrac[0.1]{mm}{h}$ & 7.4\%
& 61\% \\
\hline
2 & Continuous light rain and higher & $ \CRF \ge \unitfrac[1]{mm}{h}$ & 2.9\%
& 43\% \\
\hline
3 & Moderate rain and higher & $ \CRF \ge \unitfrac[2.5]{mm}{h}$ & 1.2\% & 34\% \\ 
\hline
\end{tabular}

\end{table}

Knowing $\widetilde{M}_{(k-1,k-s)}$, the classifier $m$ will estimate the probability $\P_{i,j,k}^m$ that the cumulative rainfall $\CRF_{i,j,k+p}$ belongs to $\C_m$:
\begin{equation}
    \P_{i,j,k}^m = \P_m(\CRF_{i,j,k+p} \in \C_m \mid \widetilde{M}_{(k-1,k-s)}) = \P_m(\CRF_{i,j,k+p} \geq L_m \mid \widetilde{M}_{(k-1,k-s)})
\end{equation}
with $\P_{i,j,k}^m \in [0,1]$ reaching 1 if $\CRF_{i,j,k+p}$ surely belongs to class $\C_m$. Ultimately, all values in the sequence of probabilities $(\P_{i,j,k}^m)_{m\in\{1,\cdots,N_L\}}$ that are above 0.5 mark that data as belonging to $\C_m$. 
When no classifier satisfies $\P_{i,j,k}^m \geq 0.5$, no rain is predicted.

\section{Data}
\label{section_data}

MeteoNet~\cite{meteonet} is a M\'et\'eo-France project gathering meteorological data on the French territory. 
Every data type available spans from 2016 to 2018 on two areas of 500 {km} $\times$ 500 {km} each, framing the northwest and southeast parts of the French metropolis. 
This paper focuses on rain radar and wind data in the northwest area.

\subsection{Rain Data}
\label{subsection_rain}

\subsubsection{Presentation of the Meteonet Rain Radar Images and Definition of the Study Area}

The rain data in the northwest part of France provided by MeteoNet is the cumulative rainfall over time steps of 5 min. The acquisition of the data is made using M\'et\'eo-France Doppler radar network: each radar scans the sky to build a 3D reflectivity map, the different maps are then checked by  M\'et\'eo-France to remove meteorological artifacts and to obtain MeteoNet rainfall data. The spatial resolution of the data is 0.01 degrees (roughly 1 {km} $\times$ 1.2 {km}). 
More information can be found in~\cite{Meteofrance_radar_newtork} about the M\'et\'eo-France radar network and in~\cite{these_assimilation_variationnelle} about the measurement of rainfall.

The data presented in MeteoNet are images of size $565\times784$ pixels, each pixel's value being the $\CRF$ over 5 min. 
These images are often referred to as rainfall maps in this paper (see Figure~\ref{fig:rain_sample} for an example).

\begin{figure}[H]
\includegraphics[width=12 cm]{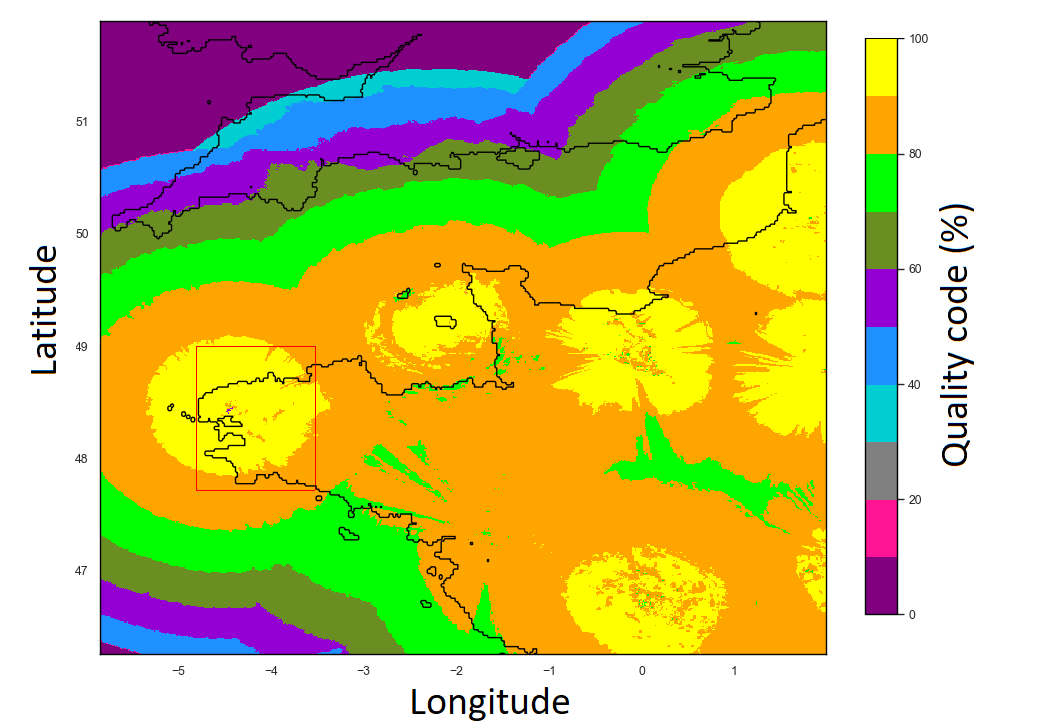}
\caption{Mean over three years (2016 to 2018) of the quality code, a score quantifying the acquisition of the rainfall data, in the northwest of France with the study area framed in red. Missing data were set to 0.}\label{fig:quality_code}
\end{figure}

The aim is to predict the rainfall at the scale of a French department hence the study area has been restricted to $128\times128$ pixels (roughly 100 {km} $\times$ 150 {km}). 
However, as the quality of the acquisition is not uniform across the territory, MeteoNet provides a rain radar quality code data (spanning from 0\% to 100\%) to quantify the quality of the acquisition on each pixel (see Figure~\ref{fig:quality_code}). The department of Finist\`ere is mainly inland and has an overall quality code score over 80\% hence the study area has been centered on the city of Brest.

\subsubsection{Definition and Distribution of Rainfall Classes in the Training Base}

Similarly to the work in ~\cite{GoogleArticle}, the three classes are defined using the following thresholds: $L_1$ = 0.1 {mm}/{h}, $L_2$ = 1 {mm}/{h}, and $L_3$ = 2.5 {mm}/{h}. 
In the following, we define three classifiers ($N_L=3$) associated with these thresholds. Classifier $m$ ($m \in \{1,2,3\}$) aims at estimating the probability of each pixel of an image to belong to the class $C_m$, with $\C_m =\{\CRF_{i,j,k} \ge \th_m\}$.

The definition of the classes is summarized in the first three columns of Table~\ref{tab:Table summary of classes definition}. Note that the scale is in millimeters of rain per hour, whereas MeteoNet provides the cumulative rainfall over 5 min, and therefore a factor of $1/12$ is applied.

Now, we further detail the distribution of the classes across the database by assessing, for each class, the proportion of $\CRF_{i,j,k}$ exceeding its threshold. To calculate these percentages, only data of the training set are considered (see Section~\ref{Splitting_the_dataset} for a definition of the training set). The results are presented in the column ``Pixels by class (\%)'' of Table~\ref{tab:Table summary of classes definition}.
One can infer from this table that the percentage of pixels corresponding to ``no rain'' ($\CRF < 0.1$) is 92.6\%. Those pixels are highly dominant, highlighting the scarcity of rain events.

\begin{figure}[H]
\includegraphics[width=12 cm]{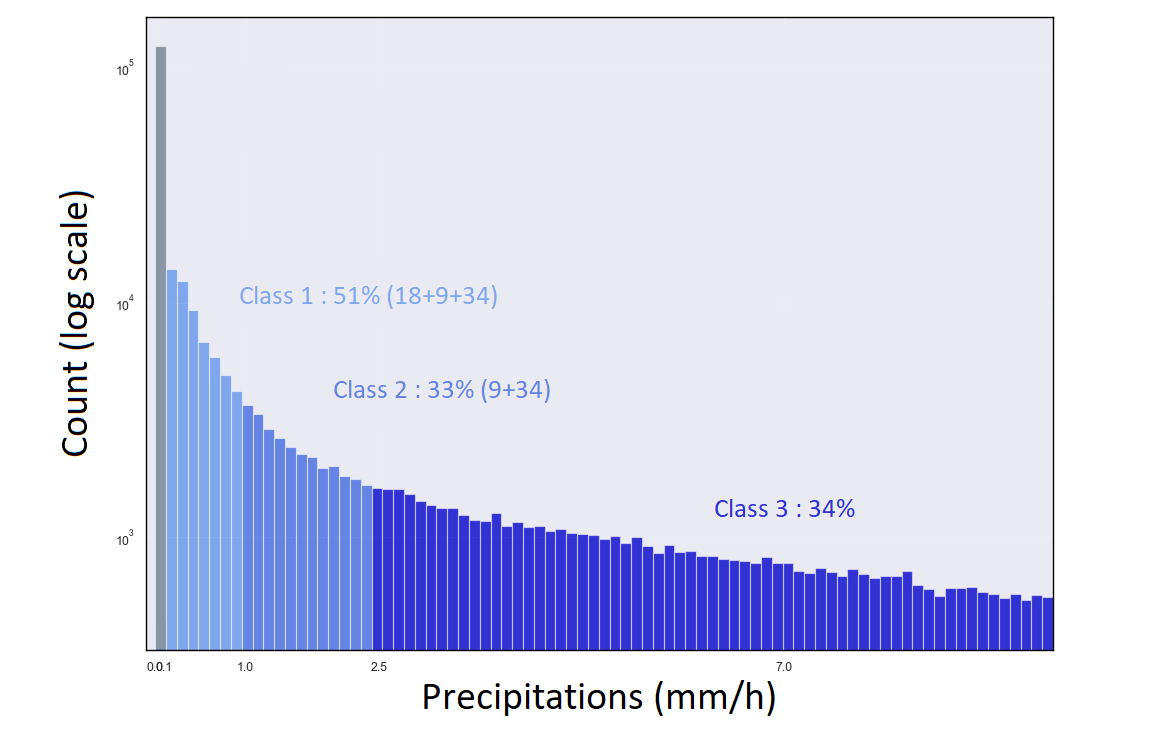}
\caption{\textls[-15]{Histogram of maximal $\CRF$ over the study area calculated on the training base and represented in log scale. The percentages correspond to the proportion of maxima belonging to each class.}}\label{fig:rain_histogram}
\end{figure}

Second, to evaluate the distribution of classes across rainfall maps (see Figure~\ref{fig:rain_sample} for an example of rainfall map), the histogram of the maximum $\CRF$ of each rainfall map restricted to the study area was calculated and is presented in Figure~\ref{fig:rain_histogram}.

Similarly to Table~\ref{tab:Table summary of classes definition}, only the data of the training base were considered. This histogram shows that data above 2.5 {mm}/{h} are present and evenly distributed among the rainfall maps: even if they only account for 1.2\% of the total proportion of the dataset, more than 30\% of rainfall maps contain at least one pixel belonging to this class (see the last column of Table~\ref{tab:Table summary of classes definition}). It is therefore likely that the data of this class form small patches distributed across the images. This phenomenon and rain scarcity are major problems in rain nowcasting; because of them, the adjustment of the weights of the neural network during the training phase is unstable for the classes of heavy rain. Therefore, higher classes are more difficult to predict. This problem is tackled in Section~\ref{subsection_oversampling}.

\subsection{Wind Data}
\label{subsection_wind}

\subsubsection{Presentation of the Meteonet Wind Data}

MeteoNet provides the weather forecasts produced by two Meteo-France weather forecast models: AROME and ARPEGE. Because it provides a better precision in both time and space, only AROME data were used. From 2016 to 2018, AROME was run every day at midnight, forecasting wind velocity and direction for every hour of the day. The wind-related data available are $U$ component (wind velocity vector component from west to east in {m}/{s}) and $V$ component (wind velocity vector components from south to north in {m}/{s}). These forecasts are made with a spatial resolution of 0.025 degrees ($\approx$1 {km}) at 10 m above the ground. The data presented in MeteoNet is equivalent to images of size $227\times315$ pixels, each pixel's value is the wind velocity at a given time. In the following, those images are referred to as wind maps (see Figure~\ref{fig:wind_sample} for an example). To fit the mesh of rainfalls, AROME data were linearly interpolated on both space and time.

\subsubsection{Distribution of the Wind Data in the Training Base}

\begin{figure}[H]

\includegraphics[width=13 cm]{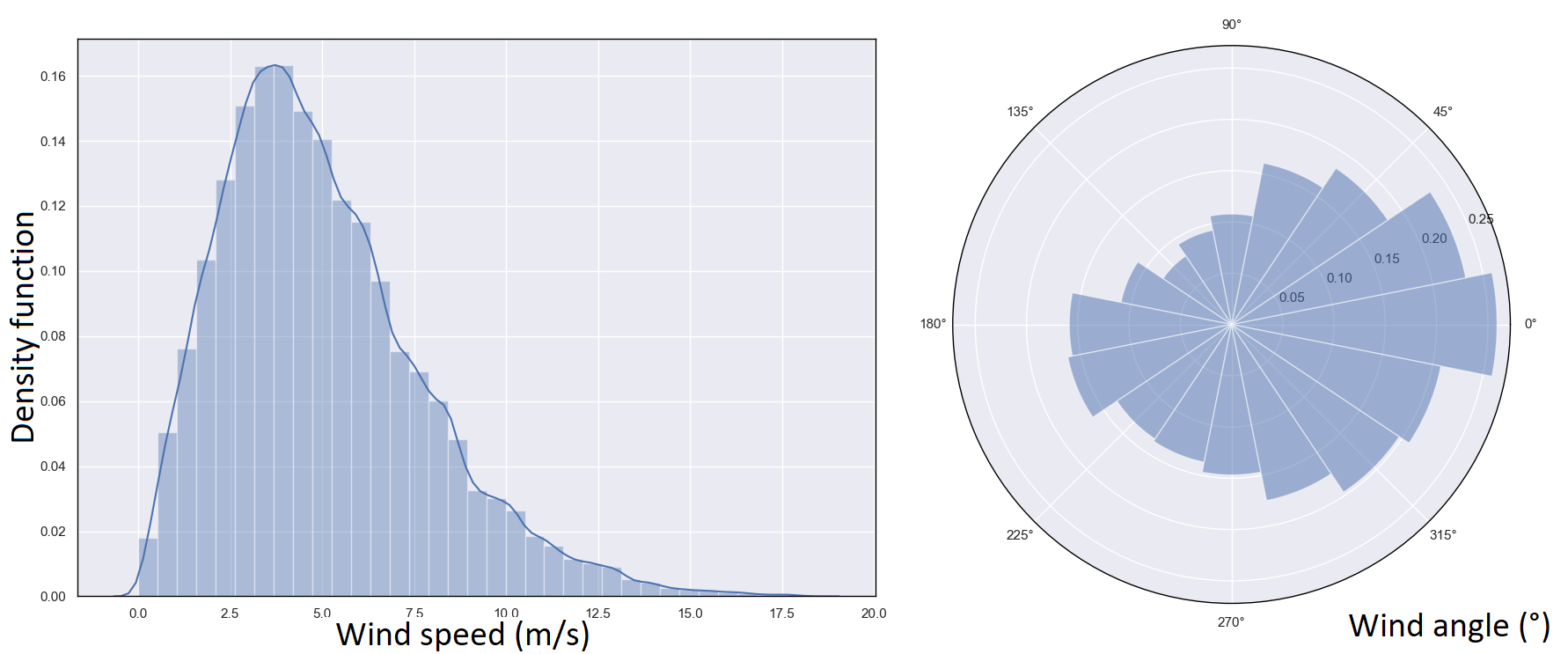}
\caption{Normalized 
 histograms and density functions of the mean wind velocity and the mean wind direction across wind maps of the training base. Left: density function (blue curve) of wind velocity norm in {m}/{s} with the normalized histogram. Right: normalized histogram of wind direction in degrees.}\label{fig:wind figures}
\end{figure}

\textls[-15]{Figure~\ref{fig:wind figures} represents the histograms of the mean wind speed and the mean wind direction across wind maps. For the calculation, only data of the training base (see Section~\ref{Splitting_the_dataset} for definition) were considered. The wind speed distribution is similar to a Gamma distribution and as expected the wind direction is mainly directed from the ocean to the shore.}

\section{Method}
\label{section_proposed_approach}

\subsection{Definition of the Datasets}
\label{Splitting_the_dataset}

The whole database is split into training, validation, and test sets. The years 2016 and 2017 are used for the training set. 
For the year 2018, one week out of two is used in the validation set and the other one in the test set. 
Before splitting, one hour of data is removed at each cut between two consecutive weeks to prevent data leakage~\cite{kaufman2012leakage}.
The splitting process is done on the whole year to assess the seasonal effects of the prediction in both validation and test sets. The training set is used to optimize the so-called trainable parameter of the neural network (see Section~\ref{subsection_training}), the validation set is used to tune the hyperparameters (see Table~\ref{tab:Table of hyper-parameters}), and the test set is finally used to estimate the final scores presented in Section~\ref{section_results}.

\nointerlineskip
\begin{table}[h]
\caption{\label{tab:Table of hyper-parameters}Table of hyperparameters used to train the networks.}
\begin{tabular}{  | c | c | c| c | c | c | } 
\hline
Epochs & Learning Rate  & Batch Size & Oversampling (\%) & Regularization & Gradient Clipping \\ 
\hline
4 and under & 0.0008 & 256 & 0.9 & $10^{-5}$ & 0.1 \\ 
\hline
Above 4 & 0.0001 & 256 & 0.9 & $5\times10^{-5}$ & 0.1 \\ 
\hline
\end{tabular}

\end{table}

The inputs of the network are sequences of MeteoNet images: 12 images collected five minutes apart over an hour are concatenated to create an input. Because we use three types of images ($\CRF$, $U$, and $V$), the dimension of the inputs is $36\times128\times128$ (12 rainfall maps, 12 wind maps $U$, 12 wind maps $V$). In the formalism defined in Section~\ref{section_problem_statement} $s=36$.

Each input sequence is associated with its prediction target which is the rainfall map $p$ time steps after the last image of the input sequence thresholded based on the $N_L=3$ different thresholds. 
The dimension of the target is $3\times128\times128$. The channel $m \in \{1,\cdots, 3\}$ is composed of binary values equal to 1 if $CRF_{i,j,k+p}$ belongs to class $m$, and 0 otherwise. These class maps are noted ${\T}_{i,j,k}^m$.

\begin{figure}[H]
\includegraphics[width=13 cm]{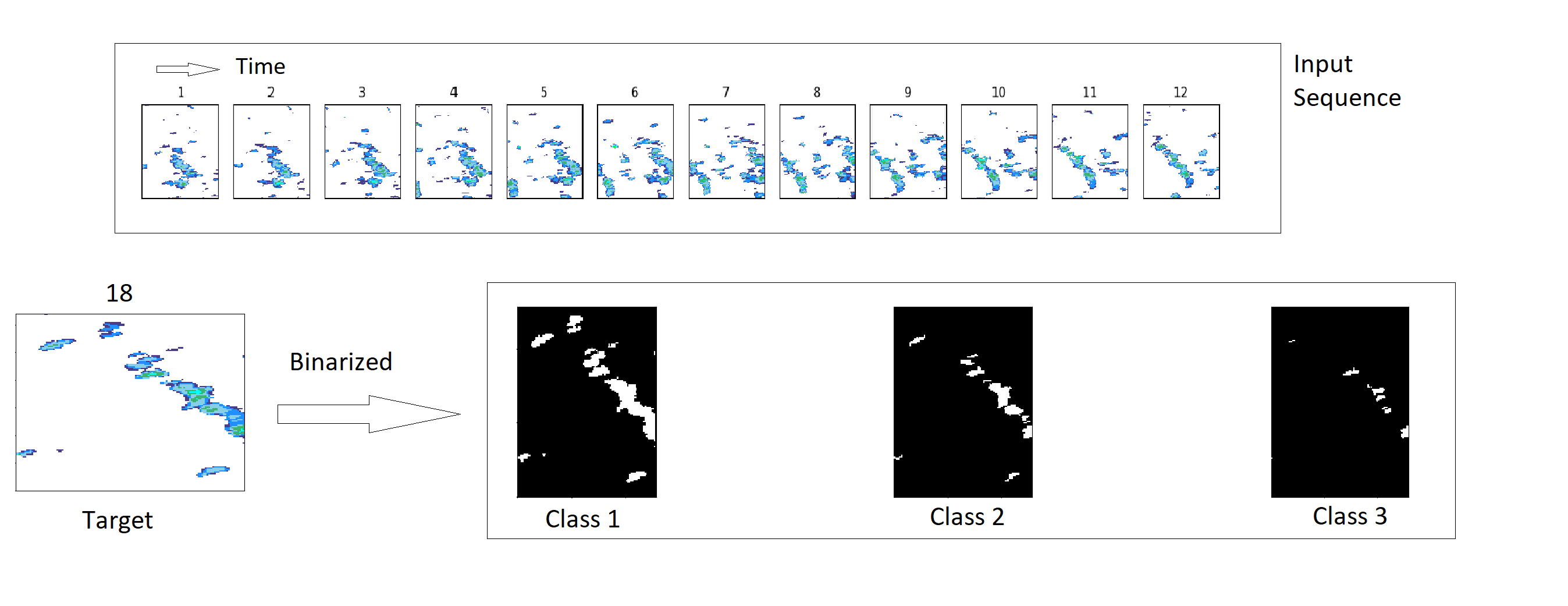}
\caption{An example of the rain channels of an input sequence starting on the 2016-01-07 at 07:00:00. On the top is represented the input sequence: 12 time steps corresponding to the rain radar images collected five minutes apart over an hour. On the bottom are represented on the left side the rain radar image collected 30 min after the last one of the input sequence, and on the right side the target (the same image thresholded to the 3 classes).}\label{fig:input sequence}
\end{figure}

An example of an input sequence and its target is given in Figure~\ref{fig:input sequence}.

If the input sequence or its target contains undefined data (due to a problem of acquisition), or if the last image of the input sequence does not contain any rain (the sky is completely clear), the sequence is set aside and will not be considered.
Each input corresponds to a distinct hour: there is no overlapping between the different inputs but note that overlapping can be an option to increase the size of the training set even if it can result in overfitting. 
It has not been used here as the actual training set contains 16,837 sequences which are considered to be large enough.
The validation set contains 4293 sequences and the test set contains 4150 sequences.
\subsubsection{Dealing with Rain Scarcity: Oversampling}
\label{subsection_oversampling}

Oversampling consists of selecting a subset of sequences of the training base and duplicating them so that they appear several times in each epoch of the training phase (see Section~\ref{subsection_training} for details on epochs and the training phase).

The main issue in rain nowcasting is to tackle rain scarcity that causes imbalanced classes. Indeed, in the training base 92.6\% of the pixels does not have rain at all (see Table~\ref{tab:Table summary of classes definition}). Therefore, the last class is the most underrepresented in the dataset and thus it will be the most difficult to predict. An oversampling procedure is thus proposed to balance this underrepresentation.
Note that the validation and test sets are left untouched to consistently represent the reality during the evaluation of the performance.

Currently, sequences whose target contains an instance of the last class represent roughly one-third of the training base (see Figure~\ref{fig:rain_histogram} and Table~\ref{tab:Table summary of classes definition}). These sequences are duplicated until their proportion in the training set reaches a chosen parameter $\ov \in [0,1]$. In practice, this parameter is chosen to be greater than the original proportion of the last class ($\ov>34\%$ in this case).

It is worth noting that the oversampling is acting image-wise and does not compensate for the unbalanced representation of class between pixels in each image. The impact and tuning of the parameter $\ov$ are discussed in Section~\ref{tuning_oversampling}.

\subsubsection{Data Normalization}
\label{subsection_normalization}
Data used as an input to train and validate the neural network are first normalized.
The normalization procedure for the rain is the following. After computing the maximum cumulative rainfall over the training dataset, $\max(\CRF)$, the following transformation is applied to each data: 
\begin{eqnarray}
CRF_{i,j,k} \leftarrow \frac{\log(1+\CRF_{i,j,k})}{\log(1+\max(\CRF))}
\end{eqnarray} 

This invertible normalization function brings the dataset into the $[0,1]$ range while spreading out the values closest to 0.

As for wind data, considering that $U$ and $V$ follow a Gaussian distribution, with \( \mu \) and \( \sigma \), respectively, the mean and the standard deviation of wind over the overall training set, we apply 
\begin{eqnarray}
U_{i,j,k} \leftarrow \frac{U_{i,j,k}-\mu_U}{\sigma_U}\\ 
V_{i,j,k} \leftarrow \frac{V_{i,j,k}-\mu_V}{\sigma_V} 
\end{eqnarray} 

\subsection{Network Architecture}
\label{subsection_architecture}
\begin{figure}[H]
\includegraphics[width=10.5 cm]{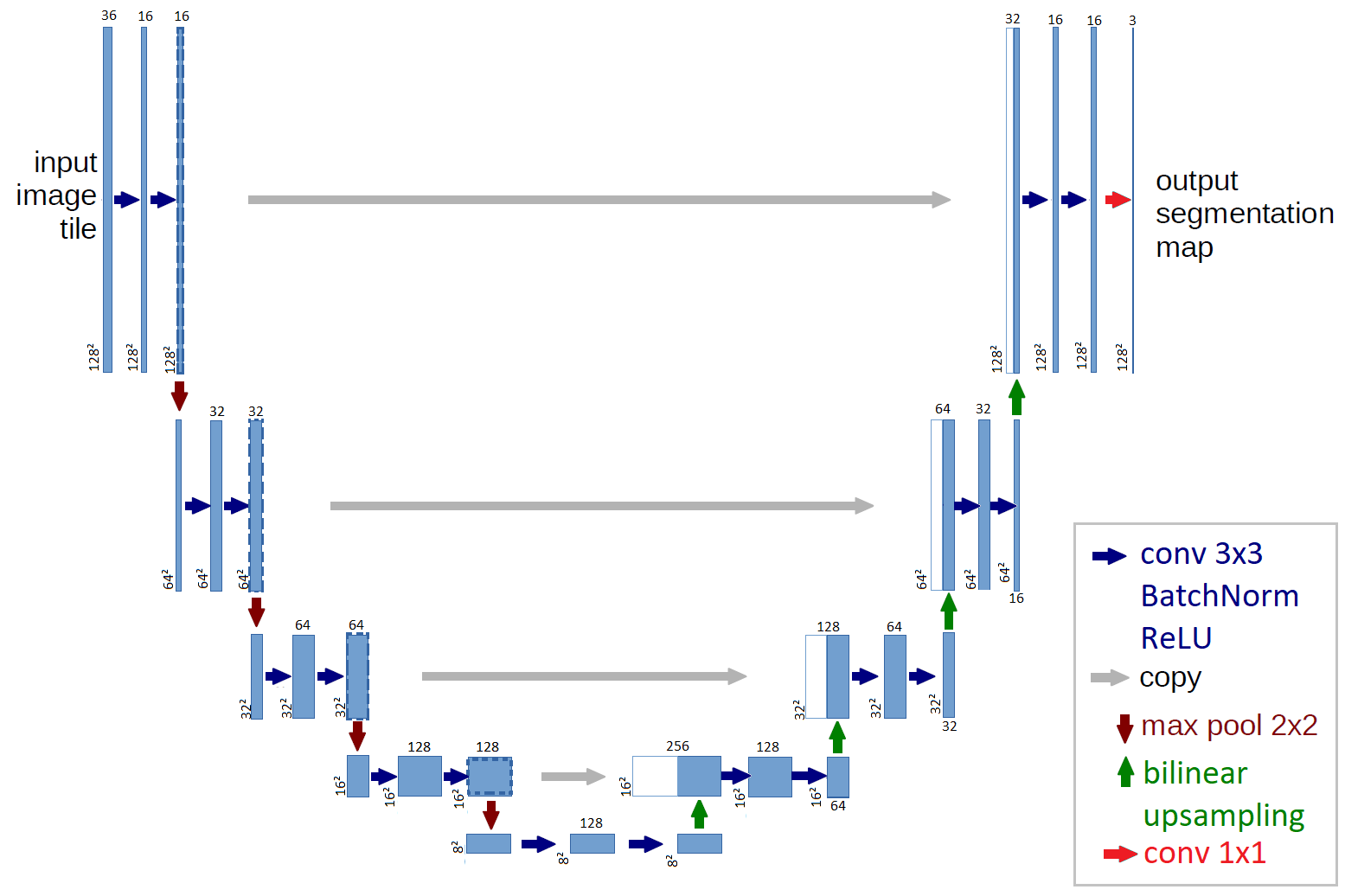}
\caption{Model architecture.}\label{fig:model}
\end{figure}

A convolutional neural network (CNN) is a feedforward neural network stacking several layers: each layer uses the output of the previous layer to calculate its own output, it is a well-established method in computer vision~\cite{Goodfellow-et-al-2016-cnn}. We decided to use a specific type of CNN, the U-Net architecture~\cite{UNET}, due to its successes in image segmentation. We chose to perform a temporal embedding through a convolutional model, rather than a Long Short-Term Memory (LSTM) or other recurrent architectures used in other studies (such as~\cite{Shi2015,Shi2017}), given that the phenomenon to be predicted is considered to have no memory (also called Markovian process). However, the inclusion of previous time steps remains warranted: the full state of the system is not observed, and the temporal coherence of the time series constraints our prediction to better fit the real rainfall trajectory. The details of the selected architecture are presented in Figure~\ref{fig:model}.

Like any U-Net network, the architecture is composed of a decreasing path also known as the encoder and an increasing path also known as the decoder. The encoding path starts with two convolutional layers. Then, it is composed of four consecutive cells, each being a succession of a max-pooling layer (red arrows in Figure~\ref{fig:model}, detailed hereafter) followed by two convolutional layers (blue arrows in Figure~\ref{fig:model}). Note that each convolutional layer used in this architecture is followed by a Batch-norm~\cite{ioffe2015batch} layer and a rectifier linear unit  (ReLU)~\cite{glorot2011deep} (Batch-norm and ReLU are detailed further down). At the bottom of the network, two convolutional layers are applied. Then the decoding path is composed of four consecutive cells each being a succession of a bilinear upsampling layer (green arrow in Figure~\ref{fig:model}, detailed hereafter) followed by two convolutional layers. Finally, a $1\times1$ convolutional layer (red arrow in Figure~\ref{fig:model}) combined with an activation function maps the output of the last cell to the segmentation map. The operation and the aim of each layer are now detailed.

\subsubsection{Convolutional Layers}
Convolutional layers perform a convolution by a kernel of $3\times3$, a padding of 1 is used to preserve the input size. The parameters of the convolutions are to be learned during the training phase. Each convolutional layer in this architecture is followed by a Batch-norm and a ReLU layer.

A Batch-norm layer re-centers and re-scales its inputs to ensure that the mean is close to 0 and the standard deviation is close to 1. Batch-norm helps the network to train faster and to be more stable~\cite{ioffe2015batch}. For an input batch Batch, the output is $y=\frac{E[Batch]}{\sqrt{V[Batch]+\epsilon}}\gamma + \beta$, where $\gamma$ and $\beta$ are trainable parameters, $E[\cdot]$ is the average, and $V[\cdot]$ is the variance of the input batch. In our architecture, the constant $\epsilon$ is set to $10^{-5}$.
 
A ReLU layer, standing for rectifier linear unit, applies the following nonlinear function \(f: x\in \mathbb{R} \mapsto \max(0,x) \). Adding nonlinearities enables the network to model non-linear relations between the input and output images.

\subsubsection{Image Sample}
To upsample or subsample the images, two types of layers are considered.

\textls[-15]{The Max-pooling layer is used to reduce the image feature sizes in the encoding part. It browses the input with a $2\times2$ filter and maps each patch to its maximum. It reduces the size of the image by a factor  between each level of the encoding path
. It also contributes to prevent overfitting by reducing the number of parameters to be optimized during the training.}

The bilinear upsampling layer is used to increase the image feature sizes in the decoding part. It performs a bilinear interpolation of the input resulting in the size of the output being twice the one of the input. 

\subsubsection{Skip Connections}
Skip connections (in gray in Figure~\ref{fig:model}) are the trademarks of the U-Net architecture. The output of an encoding cell is stacked to the output of a decoding cell of the same dimension and the stacking is used as input for the next decoding cell. Therefore, skip connections spread some information from the encoding path to the decoding path and thus help to prevent the vanishing gradient problem~\cite{drozdzal2016importance} and allow to prevent some small scale features in the encoding path.

\subsubsection{Output Layer}
 
The final layer is a convolutional layer with a $1\times1$ kernel. The dimension of its output is $3\times128\times128$, there is one channel for each class. For a given point we define the score \({s}_m\) as the output of channel $m$ ($m \in \{1,2,3\}$). This output ${s}_m$ is then transformed using the sigmoid function to obtain the probability $P_{i,j,k}^m$: 
\begin{equation}
    P_{i,j,k}^m = \frac{1}{1+ e^{s_m}}
\end{equation}
where $P_{i,j,k}^m$ is defined in Section~\ref{section_problem_statement}. Note that, following the definition of the classes in Table~\ref{tab:Table summary of classes definition}, one point can belong to several classes.

Finally, the output is said to belong to the class $m$ if $P_{i,j,k}^m \geq \frac12$. 

\subsection{Network Training}
\label{subsection_training}

We call $\theta$ the vector of length $N_\theta$ containing the trainable parameters (also name weights) that are to be determined through the training procedure. 
The training process consists of splitting the training dataset into several batches, inputting successively the batches into the network, calculating the distance between the predictions and the targets via a loss function, and finally, based on the calculated loss, updating the network weights using an optimization algorithm. The training procedure is repeated during several epochs (one epoch being achieved when the entire training set has gone through the network) and aims at minimizing the loss function.

For a given input sequence $\widetilde{M}_{(k-1,k-s)}$, we define the binary cross-entropy loss function~\cite{Goodfellow-et-al-2016-cnn-3} $Loss$ comparing the output $P_{i,j,k}^m$ to its target $\T_{i,j,k}^m$:
\begin{equation}
Loss(\theta) = - \frac{1}{N_L} \sum\limits_{m=1}^{N_L} \left( \T_{i,j,k}^m \log(\P_{i,j,k}^m) + (1-\T_{i,j,k}^m) \log (1-\P_{i,j,k}^m) \right )
\end{equation}

This loss is averaged across the batch, then a regularization term is added: 
\begin{equation}
Regularization(\theta) = \frac{\delta}{N_\theta} \sum\limits_{l=1}^{N_\theta} \theta_l^2
\end{equation}

The loss function minimizes the discrepancy between the targeted value and the predicted value, and the second term is a square regularization (also called Tikhonov or $\ell_2$-regularization) aiming at preventing overfitting and distributing the weights more evenly. The importance of this regularization in the training process is weighted by the factor $\delta$.

The optimization algorithm used is Adam~\cite{Adam} (standing for Adaptive Moment Estimation), which is a stochastic gradient descent algorithm. The recommended parameters are used: $\beta_1 = 0.9$, $\beta_2 = 0.999$ and $\epsilon = 10^{-8}$. 

Moreover, to prevent an exploding gradient, the gradient clipping technique is used. it consists of re-scaling the gradient if it becomes too large to keep it small.

The training procedure for the two neural networks is the following.
\begin{itemize}
    \item The network whose horizon time is 30 min is trained on 20 epochs. Initially, the learning rate is set to 0.0008 and after 4 epochs it is reduced to 0.0001. After epoch 13, the validation \textit{F1}-score (the \textit{F1}-score is defined in Section~\ref{subsection_scores}) is not increasing. We selected the weights optimized after epoch 13 because their \textit{F1}-score are the highest on the validation set.
    \item The network whose horizon time is 1 h is trained on 20 epochs. Initially, the learning rate is set to 0.0008 and after 4 epochs it is reduced to 0.0001. After epoch 17, the validation \textit{F1}-score is not increasing. We selected the weights of epoch 17 because their \textit{F1}-score are the highest on the validation set.
\end{itemize}

The network is particularly sensitive to hyperparameters, specifically the learning rate, the batch size, and the percentage of oversampling. The tuning of the oversampling percentage is detailed in Section~\ref{subsection_oversampling}). The other hyperparameters used to train our models are presented in Table~\ref{tab:Table of hyper-parameters}.


Neural networks were implemented and trained using PyTorch 1.5.1.
on a computer with a CPU Intel(R) Xeon(R) CPU E5-2695 v4, 2.10GHz, and a GPU PNY Tesla P100 (12GB).

For the implementation details, please refer to the code available online: some demonstration code to train the network, the weights, and an example of usage is available on the GitLab repository ({
\url{https://github.com/VincentBouget/rain-nowcasting-with-fusion-of-rainfall-and-wind-data-article}}) and archived in Zenodo ({\url{https://zenodo.org/record/4284847}}).
\subsection{Scores}
\label{subsection_scores}

Among several metrics presented in the literature~\cite{Scores_multi_label_classification, Shi2015}, the \textit{F1}-score, the Threat Score (\textit{TS}), and the \textit{BIAS} have been selected. The algorithm seems unable to predict the small scales resulting in smooth borders expressing the uncertainty of the retrieval for these features. This is expected, given that, at small scales, rainfalls are usually related to other processes than the advection of the rain cell (e.g., intensive convection) have been selected.
As our algorithm is a multi-label classification problem, each of the $N_L$ classifiers is assessed independently of the others.
For a given input sequence $\widetilde{M}_{(k-1,k-s)}$, we compare the output ($P_{i,j,k}^m$ thresholded by 0.5) to its target $\T_{i,j,k}^m$. Because it is a binary classification, four possible outcomes can be obtained: 
\begin{itemize}
\item True Positive $\mathit{TP}_{i,j,k}^m$ when the classifier rightly predict the occurrence of an event (also called hits).
\item True Negative $\mathit{TN}_{i,j,k}^m$ when the classifier rightly predict the absence of an event.
\item False Positive $\mathit{FP}_{i,j,k}^m$ when the classifier predicts the occurrence of an event that has not occurred (also called false alarm).
\item False Negative $\mathit{FN}_{i,j,k}^m$ when the classifier predicts the absence of an event that has occurred (also called missed).
\end{itemize}

On the one hand, we can define the threat score and the \textit{BIAS}: 
\begin{eqnarray}
\textit{TS}_m&=&\displaystyle\frac{\sum\limits_{i,j,k} \mathit{TP}_{i,j,k}^m}{\sum\limits_{i,j,k} \mathit{TP}_{i,j,k}^m + \mathit{FP}_{i,j,k}^m + \mathit{FN}_{i,j,k}^m} \\
\mathit{BIAS}_m&=&\displaystyle\frac{\sum\limits_{i,j,k} \mathit{TP}_{i,j,k}^m + \mathit{FP}_{i,j,k}^m}{\sum\limits_{i,j,k} \mathit{TP}_{i,j,k}^m + \mathit{FN}_{i,j,k}^m} 
\end{eqnarray} 

\textit{TS} range from 0 to 1, where 0 is the worst possible classification and 1 is a perfect classifier.
\textit{BIAS} range from 0 to $+\infty$, 1 corresponds to a non-biased classifier. A score under 1 means that the classifier underestimates the rain and a score greater than 1 means that the classifier overestimates the rain.

On the other hand, we can define the precision and the recall:
\begin{eqnarray}
\mathit{Precision}_m&\displaystyle=\frac{\sum\limits_{i,j,k} \mathit{TP}_{i,j,k}^m}{\sum\limits_{i,j,k} \mathit{TP}_{i,j,k}^m + \mathit{FP}_{i,j,k}^m}  \\ 
\mathit{Recall}_m&\displaystyle=\frac{\sum\limits_{i,j,k} \mathit{TP}_{i,j,k}^m}{\sum\limits_{i,j,k} \mathit{TP}_{i,j,k}^m + \mathit{FN}_{i,j,k}^m} \\
\end{eqnarray} 

\textls[-20]{Note that, in theory, if the classifier is predicting 0 for all the data (i.e., no rain), the Precision is not defined because its denominator is null. Nevertheless, as the simulation is done over all the samples of the validation or test dataset, this situation hardly occurs in practice.}

Based on those definitions, the $\mathit{F1}$-score, $\mathit{F1}_m$ for the classifier $m$, can be defined as the harmonic mean between the precision and the recall as
\begin{eqnarray}
\mathit{F1}_m = 2\frac{\mathit{Precision}_m\times \mathit{Recall}_m}{\mathit{Precision}_m+\mathit{Recall}_m}
\end{eqnarray} 

$\mathit{Precision}$, $\mathit{Recall}$, and $\mathit{F1}$-score range from 0 to 1, where 0 is the worst possible classification and 1 is a perfect classifier. 

All these scores will be computed on the test dataset to assess our models' performance.

\subsection{Baseline}
We briefly present the optical flow method used in Section~\ref{section_results} as a baseline.
If $I$ is a sequence of images (in our case, a succession of $\CRF$ maps), the optical flow assumes the advection of $I$ by velocity $W=(U,V)$ at pixel $(x,y)$ and time $t$:
\begin{equation}
    \frac{\partial I}{\partial t}(x,y,t) + \nabla I(x,y,t)^T W(x,y,t) = 0 \label{eq:of}
\end{equation}
where $\nabla$ is the gradient operator and $T$ the transpose operator, i.e., $\nabla I^T=\begin{pmatrix} \frac{\partial I}{\partial x} & \frac{\partial I}{\partial y}\end{pmatrix}$.
Recovering velocity $W$ from images $I$ by inverting Equation~\eqref{eq:of} is an ill-posed problem. The classic approach~\cite{horn81} is to restrict the space of solution to smooth functions using Tikhonov regularization. To estimate the velocity map at time $t$, denoted $W(.,.,t)$, the following cost-function is minimized:

\nointerlineskip
\begin{equation}
E(W(.,.,t) = \iint_\Omega \left( \frac{\partial I}{\partial t}(x,y,t) + \nabla I(x,y,t)^T W(x,y,t) \right)^2 dxdy  +\alpha \iint_\Omega\|\nabla W(x,y,t)\|^2 dxdy
\end{equation}

$\Omega$ stands for the image domain. Regularization is driven by the hyperparameter $\alpha$. The gradient is easily derived using calculus of variation.
As the cost function $E$ is convex, standard convex optimization tools can be used to obtain the solution. 
This approach is known to be limited to small displacements. 
A solution to fix this issue is to use a data assimilation approach as described in~\cite{zebiri:hal-02048500}.
Once the estimation of velocity field $\widehat{W}=(\widehat U,\widehat V)$ is computed, the last observation $I_{\mathrm{last}}$ is transported,  Equation~\eqref{eq:adI}, at the wished temporal horizon. 
The dynamics of thunderstorm cells is nonstationary, the velocity should also be transported by itself, Equation~\eqref{eq:adW}. 
Finally the following system of equations is integrated in time to the wished temporal horizon $t_h$.
\begin{eqnarray}
 \frac{\partial I}{\partial t}(x,y,t) + \nabla I(x,y,t)^T W(x,y,t) &=& 0 \quad t\in [t_0,t_h]\label{eq:adI}\\
  \frac{\partial W}{\partial t}(x,y,t) + \nabla W(x,y,t)^T W(x,y,t) &=& 0 \quad t\in [t_0,t_h] \label{eq:adW}\\
 I(x,y,t_0) &=& I_{\mathrm{last}}(x,y) \nonumber\\
  W(x,y,t_0) &=& \widehat{W}(x,y) \nonumber
\end{eqnarray}
and provide the forecast $I(t_h)$. Equations~\eqref{eq:adI} and~\eqref{eq:adW} are both approximated using an Euler and semi-Lagrangian scheme.

\section{Results}
\label{section_results}

According to the training procedure defined in Section~\ref{subsection_training}, we trained several neural networks. Using both wind maps and rainfall maps as inputs, a neural network was trained for predictions at a lead time of 30 min and another one for predictions at a lead time of 1 h. Using only rainfall maps as inputs, a neural network was trained for predictions at a lead time of 30 min and another one for predictions at a lead time of 1 h; these two neural networks provide comparison models and are used to assess the impact of wind on the forecasts.
The results are compared with the naive baseline given by the persistence model, which consists of taking the last rainfall map of an input sequence of prediction and to the optical flow approach.

\begin{figure}[H]
\includegraphics[width=10.5 cm]{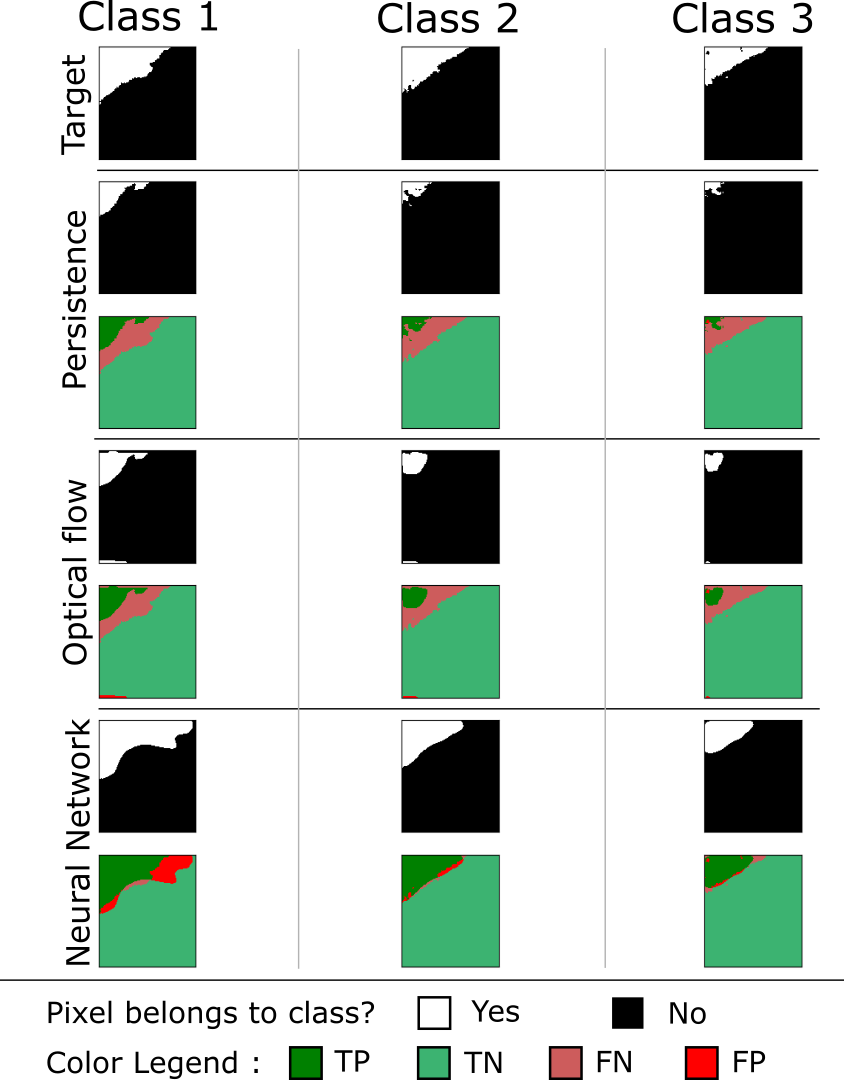}
\caption{Comparison of forecasts made by different models for a lead time of 30 min. The four rows, respectively, correspond to the target, the persistence, the optical flow, and the neural network. For each forecast, in addition to the raw prediction, the difference to the target is given. The gap between the persistence and the target reveals the evolution of the rain field during the elapsed time (including its transport, as well as cloud seeding, evaporation, condensation, etc).}\label{fig:prediction30m}
\end{figure}

\begin{figure}[H]
\includegraphics[width=10.5 cm]{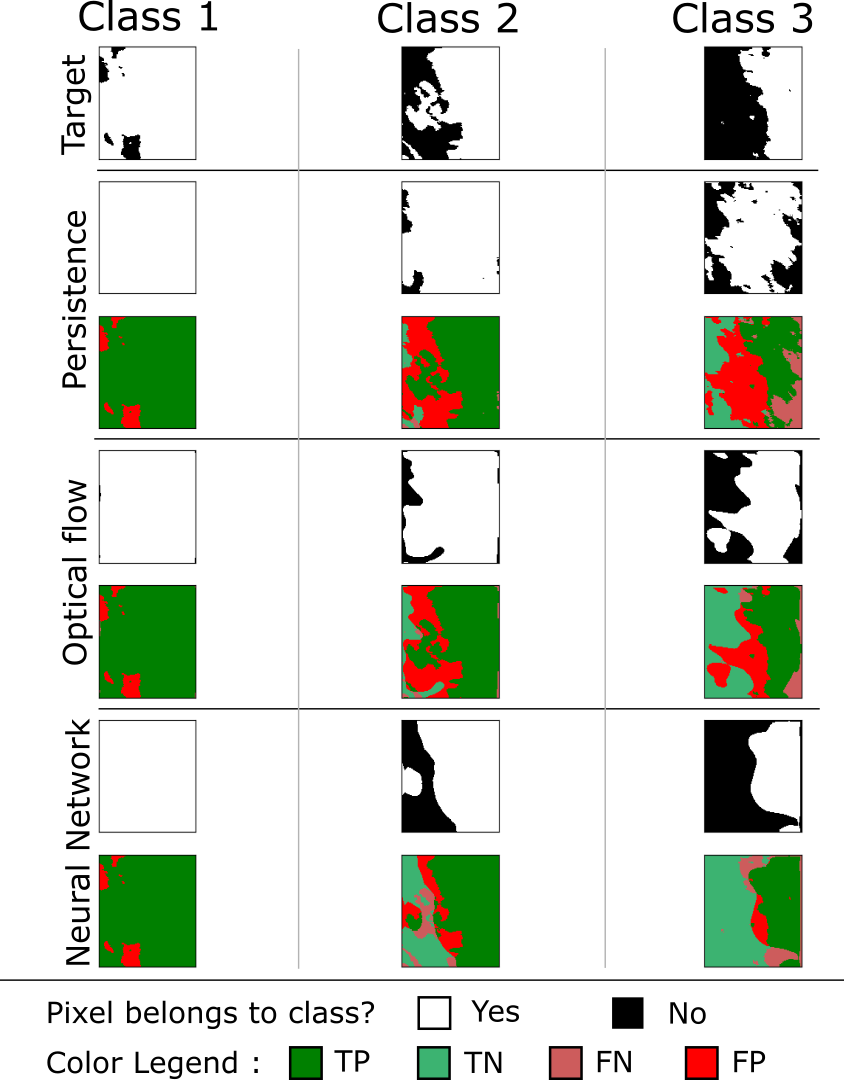}
\caption{For caption details see Figure~\ref{fig:prediction30m}. The neural network tends to smooth the rain cells and its forecast contrary to the target which is quite sparse.}\label{fig:prediction1h}
\end{figure}

Figures~\ref{fig:prediction30m} and \ref{fig:prediction1h} present two examples of prediction at 30 min made by the neural networks trained using rainfalls and wind. The forecast is compared to its target, to the persistence, and to the optical flow. The comparison shows that the network is able to model advection to be quite close to the target. 
The algorithm seems unable to predict the small scales resulting in smooth borders expressing the uncertainty of the retrieval for these features. This is expected, given that, at small scales, rainfalls are usually related to other processes than the advection of the rain cell (e.g., intensive convection). 

Using the method proposed in~\cite{Nature}, the results, presented in Table~\ref{tab:Table of results}, are calculated on the test set and 100 bootstrapped samples are used to calculate means and standard deviations of the \textit{F1}-score. 
First, it can be noticed that the neural net using both rainfalls maps and wind (denoted NN in Table~\ref{tab:Table of results}) outperforms the baselines (PER: Persistence; OF: optical flow) on both the \textit{F1}-score and the \textit{TS}. The difference is significant for the \textit{F1}-score at a lead-time of 30 min. For class 3, the 1-hour prediction presents a similar performance for the neural network and the optical flow. This is not surprising because the optical flow is sensitive to the structures' contrast and responds better when this contrast is large, which is the case of pixels belonging to class 3. 

In contrast, the bias (\textit{BIAS}) is significantly lower than 1 for the neural network, which indicates that the neural network is predicting on average less rain than observed. It is likely due to the imbalance classes within each image which is not fully compensated by the oversampling procedure. This is confirmed by the fact that class 3, which is the most underrepresented in the training set, has the lowest bias.
The performances of the neural net using only the rainfalls images in the input (denoted NN/R) are also reported in Table~\ref{tab:Table of results}. It can be seen that the addition of the wind in input provides a significant improvement in all cases (with a maximum of 10\% for class 3 at a lead time of 30 min) for the \textit{F1}-score. The improvement is even greater for higher classes which are the most difficult to predict. However, the difference is less significant for a lead-time of 1 h. Quite interestingly, the addition of the wind in the input is also able to reduce the bias of the neural network, suggesting that having adequate predictors is a way to both improve the skill and reduce the bias.

    
    
    

\begin{table}[H]
  \caption{\label{tab:Table of results}Comparison of models' results on \textit{F1}, \textit{TS}, and \textit{BIAS} scores (mean $\pm$ standard deviation) for the three classes calculated on the test set (bold numbers denote the best score).}
  
\resizebox{\textwidth}{!}{\begin{tabular}{cccccccccccllllll}
\cline{1-13}
\multirow{2}{*}{\begin{tabular}[c]{@{}c@{}}Time\\ \end{tabular}} & \multirow{2}{*}{Model} & \multicolumn{3}{c}{\textit{F1}-Score (mean $\pm$ std)} &  & \multicolumn{3}{c}{\textit{TS} (mean $\pm$ std)} &  & \multicolumn{3}{c}{\textit{BIAS} (mean $\pm$ std)} &  &  &  &  \\ \cline{3-5} \cline{7-9} \cline{11-13}
 &  & Class 1 & Class 2 & Class 3 &  & Class 1 & Class 2 & Class 3 &  & Class 1 & \multicolumn{1}{c}{Class 2} & \multicolumn{1}{c}{Class 3} \\ \cline{1-13}
\multirow{4}{*}{30 min} & \textbf{PER} & 0.56 $\pm$ 0.07 & 0.38 $\pm$ 0.10 & 0.23 $\pm$ 0.10 & & 0.41 $\pm$ 0.06 & 0.25 $\pm$ 0.07 & 0.13 $\pm$ 0.07 & & {\bf 1.01} $\pm$ 0.06 & {\bf 1.00} $\pm$ 0.11 & {\bf 0.96} $\pm$ 0.17 \\
 & \textbf{OF} & 0.59 $\pm$ 0.11 & 0.49 $\pm$ 0.13 & 0.37 $\pm$ 0.15 & & 0.44 $\pm$ 0.14 & 0.33 $\pm$ 0.10 & 0.20 $\pm$ 0.07 & & 1.03 $\pm$ 0.12 & 0.91 $\pm$ 0.16 & 0.71 $\pm$ 0.20 \\
 & \textbf{NN/R} & 0.70 $\pm$ 0.06 & 0.55 $\pm$ 0.07 & 0.36 $\pm$ 0.10 & & 0.57 $\pm$ 0.08 & 0.44 $\pm$ 0.10 & 0.27 $\pm$ 0.09 & & 0.85 $\pm$ 0.10 & 0.71 $\pm$ 0.13 & 0.54 $\pm$ 0.23 \\
 & \textbf{NN} & {\bf 0.76} $\pm$ 0.04 & {\bf 0.58} $\pm$ 0.05 & {\bf 0.46} $\pm$ 0.06 & & {\bf 0.61} $\pm$ 0.07 & {\bf 0.51} $\pm$ 0.09 & {\bf 0.35} $\pm$ 0.09 & & 0.88 $\pm$ 0.08 & 0.74 $\pm$ 0.13 & 0.59 $\pm$ 0.22 \\ \cline{1-13}
\multirow{4}{*}{60 min} & \textbf{PER} & 0.27 $\pm$ 0.06 & 0.13 $\pm$ 0.05 & 0.06 $\pm$ 0.03 & & 0.19 $\pm$ 0.08 & 0.09 $\pm$ 0.07 & 0.05 $\pm$ 0.06 & & {\bf 0.99} $\pm$ 0.06 & {\bf 0.95} $\pm$ 0.09 & {\bf 0.94} $\pm$ 0.16 \\
 & \textbf{OF} & 0.51 $\pm$ 0.06 & 0.37 $\pm$ 0.07 & 0.18 $\pm$ 0.08 & & 0.38 $\pm$ 0.16 & 0.21 $\pm$ 0.09 & 0.07 $\pm$ 0.03 & & 1.09 $\pm$ 0.10 & 0.76 $\pm$ 0.21 & 0.51 $\pm$ 0.32 \\
 & \textbf{NN/R} & 0.54 $\pm$ 0.07 & 0.34 $\pm$ 0.10 & 0.13 $\pm$ 0.08 & & 0.43 $\pm$ 0.08 & 0.25 $\pm$ 0.09 & 0.09 $\pm$ 0.05 & & 0.75 $\pm$ 0.13 & 0.61 $\pm$ 0.26 & 0.22 $\pm$ 0.19 \\
 & \textbf{NN} & {\bf 0.55} $\pm$ 0.04 & {\bf 0.41} $\pm$ 0.06 & {\bf 0.19} $\pm$ 0.05 & & {\bf 0.44} $\pm$ 0.07 & {\bf 0.27} $\pm$ 0.09 & {\bf 0.12} $\pm$ 0.06 & & 0.79 $\pm$ 0.10 & 0.68 $\pm$ 0.24 & 0.31 $\pm$ 0.21 \\ \cline{1-13}
\multicolumn{1}{l}{} & \multicolumn{1}{l}{} & \multicolumn{1}{l}{} & \multicolumn{1}{l}{} & \multicolumn{1}{l}{} & \multicolumn{1}{l}{} & \multicolumn{1}{l}{} & \multicolumn{1}{l}{} & \multicolumn{1}{l}{} & \multicolumn{1}{l}{} & \multicolumn{1}{l}{} &  &  &  &  &  & 
\end{tabular}}

\end{table}

\section{Discussion}
\label{section_discussion}

\subsection{Choice of Input Data}
\textls[-15]{One objective of this work was to show that adding relevant data as input could cause a significant improvement in nowcasting. We expect that the choice of input data can have an incidence of the improvement of the prediction skill. In this work, the 10 m wind was considered as a proxy for all the factors that could potentially influence the rain cloud formation, motion, and deformation. In particular, we did not choose the wind as the cloud level given the huge uncertainty of the cloud height and thickness~\cite{sun2007integrated}.  It has also been chosen because it is a standard product distributed in operational products and therefore it is reasonable to assume that we can estimate this parameter in a near-real-time context that could be a future application of our approach. Limiting the input factors to the wind and the radar images will limit the predictability of the algorithm. In particular, we cannot predict the formation of clouds. In future work, it could worth investigating other parameters such as 2 m air temperature, surface geopotential, or vertical profiles of geopotential and wind speeds that could help to predict other mechanisms leading to precipitations.}

\subsection{Dependency of Classes}
\label{independant_classifiers}
Following the class definition in Section~\ref{section_problem_statement}, the classes are not independent, given that a point belonging to $C_m$ would imply that it also belongs to $C_n$, $\forall 1 \le n < m$. This constraint was not explicitly imposed on the classifier, and as such it could violate this principle. In practice, we have not observed this phenomenon. Alternative modeling based on the classifier chains method~\cite{ClassifierChainMethod} would take this drawback into account.

\subsection{Tuning the Oversampling Parameter}
\label{tuning_oversampling}

\begin{figure}[H]
\includegraphics[width=10.5 cm]{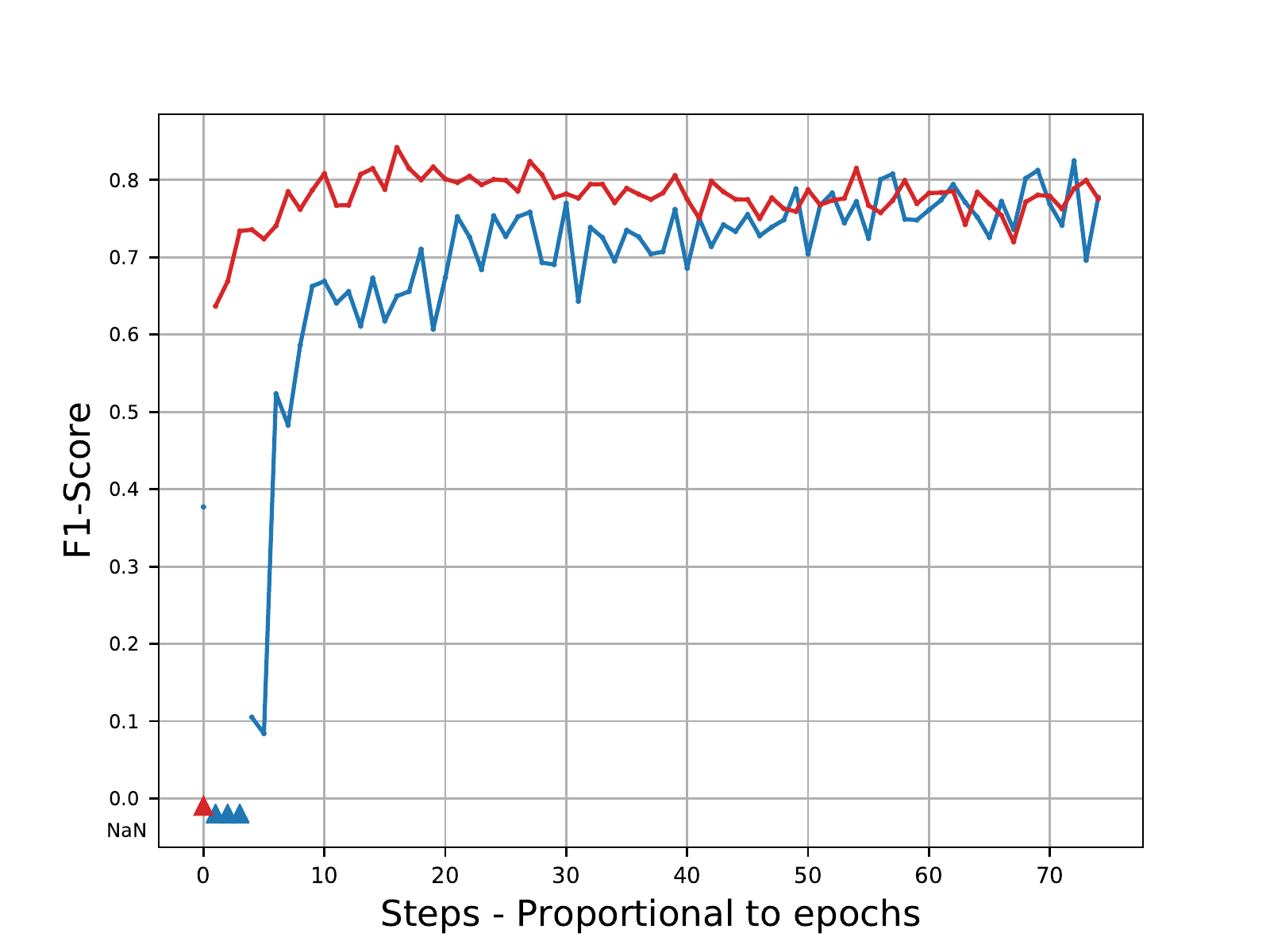}
\caption{Evolution of \textit{F1}-Score during training for Class 1 calculated on validation set. Blue curve corresponds to a training set with no oversampling ($\ov\approx0.3$) and red curve correspond to a training oversampled with $\ov=0.9$. Triangles stand for undefined \textit{F1}-score.}\label{fig:Oversample Class1 p0.8}
\end{figure}
\begin{figure}[H]
\includegraphics[width=10.5 cm]{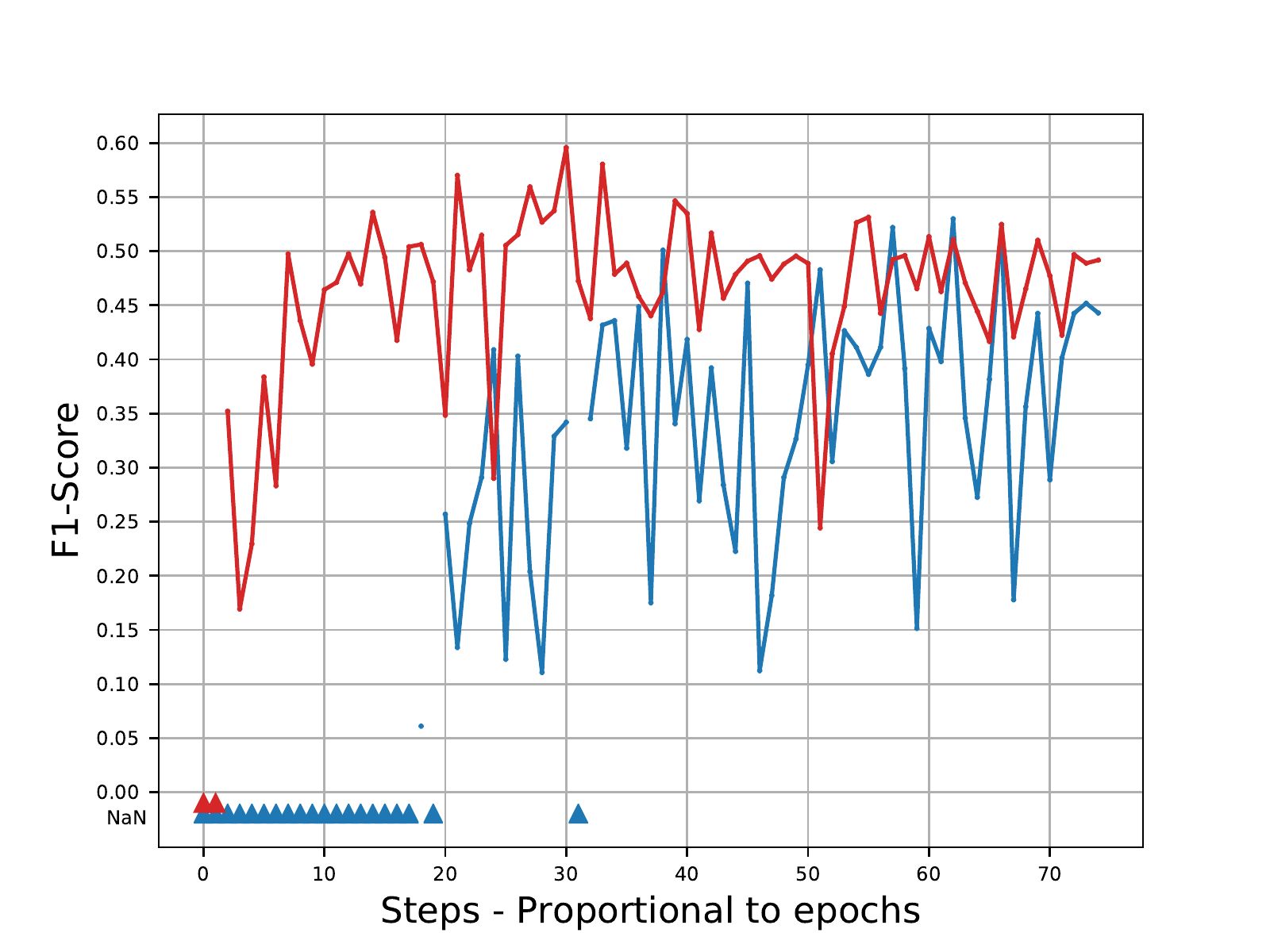}
\caption{Evolution of \textit{F1}-Score during training for Class 3 calculated on validation set. Blue curve corresponds to a training set with no oversampling ($\ov\approx0.3$) and red curve correspond to a training oversampled with $\ov=0.9$. Triangles stand for undefined \textit{F1}-score.}\label{fig:Oversample Class3 p0.9}
\end{figure}

The percentage of oversampling $\ov$ defined in Section~\ref{subsection_oversampling} is an important parameter as it will highly modify the training database; therefore, its impact will now be investigated. Several runs with different values of $\ov$ have been performed and the \textit{F1}-score calculated on the validation set are reported in 
Figures~\ref{fig:Oversample Class1 p0.8} and~\ref{fig:Oversample Class3 p0.9}. It appears on those figures that the oversampling procedure has three main advantages: the \textit{F1}-score converges faster, the results are higher for all classes, and most of all it stabilizes the training process. On raw data (without oversampling) the learning procedure is very unstable, and thus the results will poorly generalize. The oversampling procedure tackles this important issue. Based on this, an oversampling percentage of 90\% ($\ov=0.9$) is the optimal value: under this value the training phase is unstable and above this value, the network tends to overfit. The proportion of pixels before and after oversampling is compared in Table~\ref{tab:Table repartition oversampling}.

Note that, even though the oversampling percentage is defined for class 3, it also affects classes 1 and 2 as their include pixels from class 3. The stabilizing effect of the oversampling is similar to the training of the classifier for classes 1 and 2 (results on class 2 not shown).

\begin{table}[H]
\caption{\label{tab:Table repartition oversampling}Comparison of the proportion of pixels in each class before and after oversampling.}
\begin{tabular}{|c|c|c|} 
\hline
 & No oversampling  & With oversampling\\
Class number & $\ov \approx 0.3$ & $\ov = 0.9$ \\
\hline
1  & 7.4\% & 17.3\%\\ 
\hline
2 & 2.9\% & 8\%\\ 
\hline
3  & 1.2\% & 3\%\\ 
\hline
\end{tabular}

\end{table}

\subsection{Degradation of the Skill}

\begin{figure}[H]
\includegraphics[width=10.5 cm]{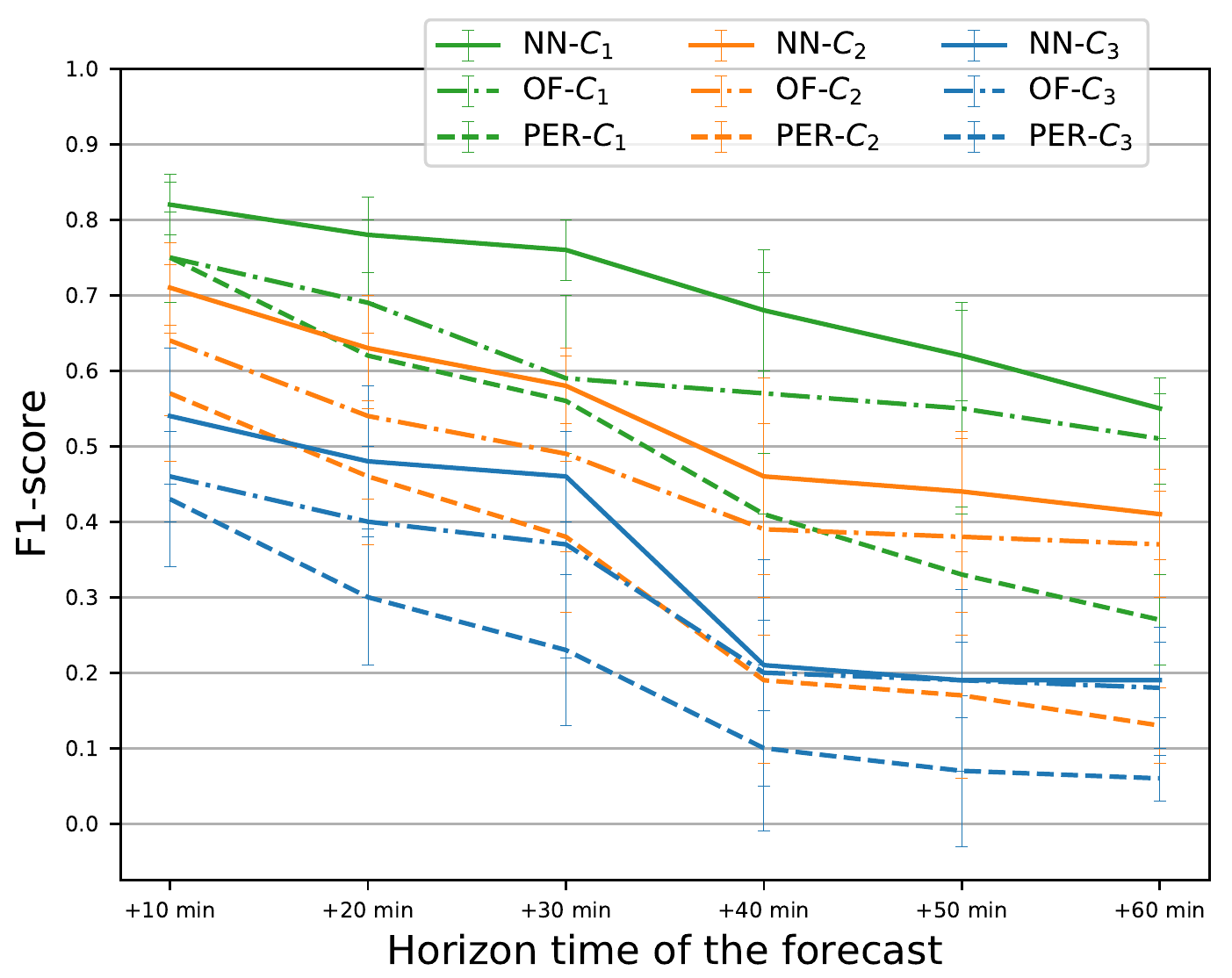}
\caption{Evolution of models' performance (in sense of \textit{F1}-score) according to the horizon forecast, from 10 min up to 1 h.}\label{fig:F1_over_time}
\end{figure}

It is expected that the prediction skill is limited depending on the lead time. For larger lead times, it becomes necessary to consider other processes such as apparition or disappearance of rain cells that can occur within or outside the region. To evaluate the degradation of the skill with respect to the lead time, a neural network (with both wind and rainfall in input) is trained for each lead time (from 10 min to one hour, every 10 min). The results of the \textit{F1}-score are presented in Figure~\ref{fig:F1_over_time}. It can be seen that the neural network prediction is consistently better than the chosen baselines (Persistence and Optical flow) for classes 1 and 2. Regarding class 3, we observe that the optical flow and the neural network reach the same minimal performance and then saturate after 40 min. It shows that, for the considered classifiers, the nowcasting skill of class 3 is limited to 30 min. Given that those classifiers rely mostly on the displacement of the rain cell, it suggests that predicting rain higher than moderate in this region would necessitate considering other physical processes than advection.
\section{Conclusion and Future Work}
\label{section_conclusion}

This work aims at studying the impact of merging rain radar images with wind forecasts to predict rainfalls in a near future. With a few meteorological parameters used as inputs, our model can forecast rainfalls with satisfactory results, outperforming the results obtained using only the radar image (without using the wind velocity).

\textls[-15]{The problem is transformed into a classification problem by defining classes corresponding to an increasing quantity of rainfall. To overcome the imbalanced distribution of these classes, we perform an oversampling of the highest class which is less frequent in the database.
The \textit{F1}-score calculated on the highest class for forecasts at a horizon time of 30 min is 45\%, our model has been compared to a basic persistence model and an approach based on optical flow and outperformed both. Furthermore, it outperforms the same architecture trained using only rainfalls up to 10\%; therefore, this paper can be considered as a proof of concept that data fusion has a significant positive impact on rain nowcasting.}

An interesting future work would be to fusion the inputs to another determining parameter, such as the orography, which could lead to overcoming some limits observe for a 1 h prediction for the class corresponding to highest rainfall.
Optical flow performances provided promising results and it would be interesting to investigate their inclusion through a scheme combining deep learning and data assimilation.

\authorcontributions{
Conceptualization, V.B., D.B., J.B., A.C., and A.F.; Data curation, V.B.; Formal analysis, V.B., D.B., J.B., A.C., and A.F.; Funding acquisition, D.B.; Methodology, V.B., D.B., J.B., A.C., and A.F.; Resources, V.B.; Software, V.B.; Supervision, D.B., J.B., A.C., and A.F.; Validation, V.B., D.B., and A.F.; Visualization, V.B.; Writing---original draft, V.B.; Writing---review and editing, D.B., J.B., A.C., and A.F.
All authors have read and agreed to the published version of the manuscript.}

\funding{J.B. have been funded by the projects REDDA (\#250711) and SFE (\#2700733) of the Norwegian Research Council. }
\acknowledgments{This project was carried out with the support of the Sorbonne Center for Artificial Intelligence (SCAI) of Sorbonne University.}

\conflictsofinterest{The authors declare no conflicts of interest.
The funders had no role in the design of the study; in the collection, analyses, or interpretation of data; in the writing of the manuscript; or in the decision to publish the results.} 

\abbreviations{The following abbreviations are used in this manuscript:\\
\noindent 
\begin{tabular}{@{}ll}
CRF & Cumulative Rainfall \\
CNN & Convolutional Neural Network \\
LSTM & Long Short-Term Memory \\
\end{tabular}}

\reftitle{References}


\begin{thebibliography}{999}

\bibitem[Schmid \em{et~al.}(2017)Schmid, Wang, and
  Harou]{WTH_rain_nowcasting_applications}
Schmid, F.; Wang, Y.; Harou, A. Nowcasting Guidelines---A Summary.
\newblock In {\em WMO---No. 1198}; World Meteorological Organization:   Geneva, Switzerland,  2017;
  Chapter~5.

\bibitem[Dixon and Wiener(1993)]{TITAN}
Dixon, M.; Wiener, G.
\newblock TITAN: Thunderstorm Identification, Tracking, Analysis, and
  Nowcasting---A Radar-based Methodology.
\newblock {\em J. Atmos. Ocean. Technol.} {\bf 1993}, {\em
  10},~785.

\bibitem[Johnson \em{et~al.}(1998{\natexlab{a}})Johnson, MacKeen, Witt,
  Mitchell, Stumpf, Eilts, and Thomas]{SCIT}
Johnson, J.T.; MacKeen, P.L.; Witt, A.; Mitchell, E.D.W.; Stumpf, G.J.; Eilts,
  M.D.; Thomas, K.W.
\newblock The Storm Cell Identification and Tracking Algorithm: An Enhanced
  WSR-88D Algorithm.
\newblock {\em Weather. Forecast.} {\bf 1998}, {\em 13},~263--276.


\bibitem[Handwerker(2002)]{TRACE3D}
Handwerker, J.
\newblock Cell tracking with TRACE3D---A new algorithm.
\newblock {\em Elsevier Atmos. Res.} {\bf 2002}, ~15--34.

\bibitem[Kyznarová and Novák(2009)]{CELLTRACK}
Kyznarová, H.; Novák, P.
\newblock {CELLTRACK---Convective cell tracking algorithm and its use for
  deriving life cycle characteristics}.
\newblock {\em Atmos. Res.} {\bf 2009}, {\em 93},~317--327.
\newblock 4th European Conference on Severe Storm.

\bibitem[Germann and Zawadzki(2002)]{MAPLE}
Germann, U.; Zawadzki, I.
\newblock Scale-Dependence of the Predictability of Precipitation from
  Continental Radar Images. Part I: Description of the Methodology.
\newblock {\em Mon. Weather. Rev.} {\bf 2002}, {\em 130},~2859--2873.

\bibitem[Bowler \em{et~al.}(2004)Bowler, Pierce, and Seed]{GANDOLF}
Bowler, N.; Pierce, C.; Seed, A.
\newblock Development of a rainfall nowcasting algorithm based on optical flow
  techniques.
\newblock {\em J. Hydrol.} {\bf 2004}, {\em 288},~74--91.

\bibitem[Shi \em{et~al.}(2015)Shi, Chen, Wang, Yeung, Wong, and Woo]{Shi2015}
Shi, X.; Chen, Z.; Wang, H.; Yeung, D.Y.; Wong, W.K.; Woo, W.C.
\newblock {Convolutional LSTM network: A machine learning approach for
  precipitation nowcasting}.
\newblock  In Proceedings of the 28th International Conference on Neural
  Information Processing Systems (NeurIPS), Montreal, QC, Canada,  7--12  December  {2015}; pp. 802--810.

\bibitem[Shi \em{et~al.}(2017)Shi, Gao, Lausen, Wang, Yeung, Wong, and
  Woo]{Shi2017}
Shi, X.; Gao, Z.; Lausen, L.; Wang, H.; Yeung, D.Y.; Wong, W.k.; Woo, W.c.
\newblock Deep Learning for Precipitation Nowcasting: A Benchmark and A New
  Model.
\newblock  In Proceedings of the 30th International Conference on Neural
  Information Processing Systems (NeurIPS),  Long Beach, CA, USA, 4--9 December 2017; pp. 5617--5627.

\bibitem[Qiu \em{et~al.}(2017)Qiu, Zhao, Zhang, Huang, Shi, Wang, and
  Chu]{China_nowcasting}
Qiu, M.; Zhao, P.; Zhang, K.; Huang, J.; Shi, X.; Wang, X.; Chu, W.
\newblock A Short-Term Rainfall Prediction Model Using Multi-task Convolutional
  Neural Networks.
\newblock  In Proceedings of the IEEE International Conference on Data Mining,  New Orleans, LA, USA, 18--21 November 2017; pp. 395--404.

\bibitem[Ayzel \em{et~al.}(2018)Ayzel, Heistermann, Sorokin, Nikitin, and
  Lukyanova]{Germany_nowcasting}
Ayzel, G.; Heistermann, M.; Sorokin, A.; Nikitin, O.; Lukyanova, O.
\newblock All convolutional neural networks for radar-based precipitation
  nowcasting.
\newblock In Proceedings of the  13th International Symposium ``Intelligent Systems
  2018'' (INTELS’18), St. Petersburg, Russia,  22--24 October  2018; pp. 186--192.

\bibitem[Hernández \em{et~al.}(2016)Hernández, Sanchez-Anguix, Julian,
  Palanca, and Duque]{Spain_nowcasting}
Hernández, E.; Sanchez-Anguix, V.; Julian, V.; Palanca, J.; Duque, N.
\newblock Rainfall Prediction: A Deep Learning Approach.
\newblock  In Proceedings of the 11th Hybrid Artificial Intelligent Systems, Seville, Spain, 18--20 April 
  2016; pp. 151--162.

\bibitem[Lebedev \em{et~al.}(2019)Lebedev, Ivashkin, Rudenko, Ganshin,
  Molchanov, Ovcharenko, Grokhovetskiy, Bushmarinov, and
  Solomentsev]{Russia_nowcasting}
Lebedev, V.; Ivashkin, V.; Rudenko, I.; Ganshin, A.; Molchanov, A.; Ovcharenko,
  S.; Grokhovetskiy, R.; Bushmarinov, I.; Solomentsev, D.
\newblock Precipitation Nowcasting with Satellite Imagery.
\newblock  In Proceedings of the 25th International Conference on Knowledge
  Discovery \& Data Mining,  Anchorage, AK, USA, 4--8 August  2019; p. 2680–2688.

\bibitem[Sato \em{et~al.}(2018)Sato, Kashima, and Yamamoto]{PredNet}
Sato, R.; Kashima, H.; Yamamoto, T.
\newblock Short-Term Precipitation Prediction with Skip-Connected PredNet.
\newblock  In Proceedings of the Internationl Conference on Artificial Neural Network and Machine
  Learning (ICANN),  Rhodes, Greece, 4--7 October  2018; pp. 373--382.

\bibitem[Lotter \em{et~al.}(2017)Lotter, Kreiman, and Cox]{PredNEt_original}
Lotter, W.; Kreiman, G.; Cox, D.
\newblock Deep Predictive Coding Networks for Video Prediction and Unsupervised
  Learning.
\newblock  In Proceedings of the International Conference on Learning Representation, Toulon, France, 24--26 April  2017.

\bibitem[Bromberg \em{et~al.}(2019)Bromberg, Gazen, Hickey, Burge, Barrington,
  and Agrawal]{GoogleArticle}
Bromberg, C.L.; Gazen, C.; Hickey, J.J.; Burge, J.; Barrington, L.; Agrawal, S.
\newblock Machine Learning for Precipitation Nowcasting from Radar Images.
\newblock  In Proceedings of the Machine Learning and the Physical Sciences Workshop at the 33rd
  Conference on Neural Information Processing Systems (NeurIPS),  Vancouver, BC, Canada, 14 December  2019; pp.
  1--4.

\bibitem[Ronneberger \em{et~al.}(2015)Ronneberger, Fischer, and Brox]{UNET}
Ronneberger, O.; Fischer, P.; Brox, T.
\newblock {U-net: Convolutional networks for biomedical image segmentation}.
\newblock  In Proceedings of the Medical Image Computing and Computer-Assisted Intervention (MICCAI), Munich, Germany, 5--9 October 
   2015; Volume 9351, pp. 234--241.

\bibitem[Larvor \em{et~al.}(2020)Larvor, Berthomier, Chabot, Le~Pape, Pradel,
  and Perez]{meteonet}
Larvor, G.; Berthomier, L.; Chabot, V.; Le~Pape, B.; Pradel, B.; Perez, L.
\newblock MeteoNet, An Open Reference Weather Dataset by Meteo-France,  2020.
\newblock  Available online:  \url{https://meteonet.umr-cnrm.fr/}   
.

\bibitem[de~Bezenac \em{et~al.}(2019)de~Bezenac, Pajot, and
  Gallinari]{de2019deep}
de~Bezenac, E.; Pajot, A.; Gallinari, P.
\newblock Deep learning for physical processes: Incorporating prior scientific
  knowledge.
\newblock {\em J. Stat. Mech. Theory Exp.} {\bf
  2019}, {\em 2019},~124009.

\bibitem[Z{\'e}biri \em{et~al.}(2019)Z{\'e}biri, B{\'e}r{\'e}ziat, Huot, and
  Herlin]{zebiri:hal-02048500}
Z{\'e}biri, A.; B{\'e}r{\'e}ziat, D.; Huot, E.; Herlin, I.
\newblock {Rain Nowcasting from Multiscale Radar Images}.
\newblock   In Proceedings of the {VISAPP 2019---14th International Conference on Computer Vision
  Theory and Applications}, Prague, Czech Republic, 25--27 February  2019; pp. 1--9.

\bibitem[Zhang \em{et~al.}(2018)Zhang, Li, Liu, et~al.]{BinaryRelevanceMethod}
Zhang, M.; Li, Y.; Liu, X.; others.
\newblock {Binary relevance for multi-label learning: an overview}.
\newblock  \emph{Front. Comput. Sci.}  \textbf{2018}, 12, 191--202.

\bibitem[Météo-France()]{Meteofrance_radar_newtork}
Météo-France.
\newblock Les Radars Météorologiques.
\newblock
 Available online:    \url{http://www.meteofrance.fr/prevoir-le-temps/observer-le-temps/moyens/les-radars-meteorologiques} (accessed on 2021-01-12) 
.

\bibitem[Mercier(2017)]{these_assimilation_variationnelle}
Mercier, F.
\newblock Assimilation Variationnelle D'observations Multi-échelles:
  Application à la Fusion de Données Hétérogènes Pour l'étude de la
  Dynamique Micro et Macrophysique des Systèmes Précipitants.
\newblock Ph.D. Thesis, Universit\'e Paris-Saclay,  Paris, France,  2017.

\bibitem[Kaufman \em{et~al.}(2012)Kaufman, Rosset, Perlich, and
  Stitelman]{kaufman2012leakage}
Kaufman, S.; Rosset, S.; Perlich, C.; Stitelman, O.
\newblock Leakage in data mining: Formulation, detection, and avoidance.
\newblock {\em ACM Trans. Knowl. Discov. Data (TKDD)} {\bf
  2012}, {\em 6},~1--21.

\bibitem[Goodfellow \em{et~al.}(2016)Goodfellow, Bengio, and
  Courville]{Goodfellow-et-al-2016-cnn}
Goodfellow, I.; Bengio, Y.; Courville, A.
\newblock {\em Deep Learning}; MIT Press:  Cambridge, MA,  USA, 2016; Chapter~9.

\bibitem[Glorot \em{et~al.}(2011)Glorot, Bordes, and Bengio]{glorot2011deep}
Glorot, X.; Bordes, A.; Bengio, Y.
\newblock Deep sparse rectifier neural networks.
\newblock  In Proceedings of the Fourteenth International Conference on Artificial
  Intelligence and Statistics,  Ft. Lauderdale, FL, USA,  11--13 April  2011; pp. 315--323.

\bibitem[Ioffe and Szegedy(2015)]{ioffe2015batch}
Ioffe, S.; Szegedy, C.
\newblock Batch normalization: Accelerating deep network training by reducing
  internal covariate shift.
\newblock  In Proceedings of the 32nd International Conference on International
  Conference on Machine Learning (ICML),  Lille, France, 6--11 July 2015; pp. 448--456.

\bibitem[Drozdzal \em{et~al.}(2016)Drozdzal, Vorontsov, Chartrand, Kadoury, and
  Pal]{drozdzal2016importance}
Drozdzal, M.; Vorontsov, E.; Chartrand, G.; Kadoury, S.; Pal, C.
\newblock The importance of skip connections in biomedical image segmentation.
  In {\em Deep Learning and Data Labeling for Medical Applications}; Springer:  Berlin/Heidelberg, Germany, 
  2016; pp. 179--187.

\bibitem[Goodfellow \em{et~al.}(2016)Goodfellow, Bengio, and
  Courville]{Goodfellow-et-al-2016-cnn-3}
Goodfellow, I.; Bengio, Y.; Courville, A.
\newblock {\em Deep Learning}; MIT Press:   Cambridge, MA, USA, 2016; Chapter~3.

\bibitem[Kingma and Lei~Ba(2015)]{Adam}
Kingma, D.; Lei~Ba, J.
\newblock Adam: A Method for Stochastic Optimization.
\newblock  In Proceedings of the 3rd International Conference for Learning Representations, San Diego, CA, USA, 7--9  May 2015.

\bibitem[Sorower(2010)]{Scores_multi_label_classification}
Sorower, M.S.
\newblock \emph{A Literature Survey on Algorithms for Multi-labelLearning};
\newblock Technical Report; Oregon State University: Corvallis, OR, USA,  2010.

\bibitem[Horn and Schunk(1981)]{horn81}
Horn, B.; Schunk, B.
\newblock Determining Optical Flow.
\newblock {\em Artif. Intell.} {\bf 1981}, {\em 17},~185--203.

\bibitem[Rajkomar \em{et~al.}(2018)Rajkomar, Oren, Chen, et~al.]{Nature}
Rajkomar, A.; Oren, E.; Chen, K.; others.
\newblock Scalable and accurate deep learning with electronic health records.
\newblock {\em npj Digit. Med.} {\bf 2018}, {\em 1},~18.

\bibitem[Sun-Mack \em{et~al.}(2007)Sun-Mack, Minnis, Chen, Gibson, Yi, Trepte,
  Wielicki, Kato, Winker, Stephens, et~al.]{sun2007integrated}
Sun-Mack, S.; Minnis, P.; Chen, Y.; Gibson, S.; Yi, Y.; Trepte, Q.; Wielicki,
  B.; Kato, S.; Winker, D.; Stephens, G.; et al.
\newblock   Integrated cloud-aerosol-radiation product using CERES, MODIS,
  CALIPSO, and CloudSat data.
\newblock  In Proceedings of the   Remote Sensing of Clouds and the Atmosphere XII. International
  Society for Optics and Photonics,  Florence, Italy, 17--19 September 2007; Volume 6745, p. 674513.

\bibitem[Read \em{et~al.}(2011)Read, Pfahringer, Holmes,
  et~al.]{ClassifierChainMethod}
Read, J.; Pfahringer, B.; Holmes, G.; Frank, E.
\newblock {Classifier chains for multi-label classification}.
\newblock  \emph{Mach. Learn.}  \textbf{2011}, 85, 333.

\end{thebibliography}
\end{document}